\documentclass[aps,superscriptaddress,notitlepage,nofootinbib]{revtex4}
\usepackage[T1]{fontenc}
\usepackage[utf8]{inputenc}
\usepackage{amsmath,amssymb,microtype,cancel,enumerate,appendix}
\usepackage[hidelinks]{hyperref}
\usepackage[x11names]{xcolor}

\DeclareMathOperator{\trace}{tr}

\makeatletter
\renewcommand{\p@subsection}{}
\makeatother

\begin{document}
\title{Gravity from Pre-geometry}

\author{Andrea Addazi}
    \email{addazi@scu.edu.cn}
    \affiliation{Center for Theoretical Physics, College of Physics Science and Technology, Sichuan University, 610065 Chengdu, China}
    \affiliation{Laboratori Nazionali di Frascati INFN, Frascati (Rome), Italy, EU}
\author{Salvatore Capozziello}
    \email{capozziello@unina.it}
    \affiliation{Dipartimento di Fisica ``E.\ Pancini'', Università di Napoli ``Federico II'', Via Cinthia 9, 80126 Napoli, Italy}
    \affiliation{Scuola Superiore Meridionale, Largo San Marcellino, 10, 80138, Napoli, Italy}
    \affiliation{INFN Sezione di Napoli, Complesso Universitario di Monte Sant'Angelo, Edificio 6, Via Cintia, 80126, Napoli, Italy}
\author{Antonino Marcianò}
    \email{marciano@fudan.edu.cn}
    \affiliation{Center for Astronomy and Astrophysics, Center for Field Theory and Particle Physics, and Department of Physics, Fudan University, Shanghai 200438, China}
    \affiliation{Laboratori Nazionali di Frascati INFN, Frascati (Rome), Italy, EU}
    \affiliation{INFN sezione di Roma ``Tor Vergata'', 00133 Rome, Italy, EU}
\author{Giuseppe Meluccio}
    \email{giuseppe.meluccio-ssm@unina.it}
    \affiliation{Scuola Superiore Meridionale, Largo San Marcellino, 10, 80138, Napoli, Italy}
    \affiliation{INFN Sezione di Napoli, Complesso Universitario di Monte Sant'Angelo, Edificio 6, Via Cintia, 80126, Napoli, Italy}
\date{\today\\,,Version accepted for publication in Classical and Quantum Gravity}

\begin{abstract}
    The gravitational interaction, as described by the Einstein--Cartan theory, is shown to emerge as the by-product of the spontaneous symmetry breaking of a gauge symmetry in a pre-geometric four-dimensional spacetime. Starting from a formulation {\it à la} Yang--Mills on an $SO(1,4)$ or $SO(3,2)$ principal bundle and not accounting for a spacetime metric, the Einstein--Hilbert action is recovered after the identification of the effective spacetime metric and spin connection for the residual $SO(1,3)$ gauge symmetry of the spontaneously broken phase -- i.e.\ the stabiliser of the $SO(1,4)$ or $SO(3,2)$ gauge group. Thus, the two fundamental tenets of General Relativity, i.e.\ diffeomorphism invariance and the equivalence principle, can arise from a more fundamental gauge principle. The two mass parameters that characterise Einstein gravity, namely the Planck mass and the cosmological constant, are likewise shown to be emergent. The phase transition from the unbroken to the spontaneously broken phase is expected to happen close to the Planck temperature. This is conjectured to be dynamically driven by a scalar field that implements a Higgs mechanism, hence providing mass to new particles, with consequences for cosmology and high-energy physics. The couplings of gravity to matter are discussed after drawing up a dictionary that interconnects pre-geometric and effective geometric quantities. In the unbroken phase where the fundamental gauge symmetry is restored, the theory is potentially power-counting renormalisable without matter, offering a novel path towards a UV completion of Einstein gravity.
\end{abstract}

\maketitle

\tableofcontents{}

\section{Introduction}
Recasting General Relativity (GR) as a gauge theory is a long-standing idea, which is rooted in the consideration of the several advantages provided by the formalism of Yang--Mills (YM) theories; among these, renormalisability is arguably the most significant, as a consistent UV completion of Einstein gravity remains elusive up to the present. Historically, approaches to reconcile the gauge principle with classical gravity have been numerous and varied, see for example the Ref.\ \cite{cabral:gauge} for a comprehensive review. Most notably, making use of the Palatini formalism, GR has been extended to its Teleparallel Equivalent formulation \cite{cai:f(T),krssak:2018ywd} and to the Einstein--Cartan theory \cite{hehl:spin}, which can be expressed as gauge theories respectively of the translation and the Poincar\'e group.\footnote{See the Ref.\ \cite{melichev:2023lwj} for recent developments on the renormalisation of Poincar\'e gauge theories.} This class of theories attempts to embed the gauge principle into GR by reinterpreting its geometric formalism, while leaving the fundamental dynamics of gravity unquestioned or only slightly extended (for instance to accommodate torsion). In general, equivalent representations of gravity can be discussed in the framework of metric-affine geometries, pointing out the basic concepts from which these theories originate \cite{capozziello:2022zzh}.\\

A radically different perspective is adopted by formulations that assume the gauge principle to be an even more fundamental aspect. These approaches seek to recover the Einstein--Hilbert (EH) action and the basic principles of GR as emergent constructions out of a more primitive action {\it à la} Yang--Mills: instead of implementing the gauge principle into GR, the latter is derived from a gauge theory which, in principle, is unrelated to gravity. This strategy has the advantage that the YM formalism can be applied from scratch, but faces the difficulty of properly reconstructing classical gravity with all of its peculiarities that make it a markedly different theory than YM ones. Important attempts in this direction were made in 1977 by {\it MacDowell} \& {\it Mansouri} \cite{macdowell:unified}, who devised a unified geometric formulation of gravitation and Supergravity, and independently by {\it Chamseddine} \& {\it West} \cite{chamseddine:supergravity}, with subsequent improvements by {\it Stelle} \& {\it West} \cite{stelle:de-sitter} in 1979 and by {\it Wilczek} \cite{wilczek:gauge} in 1998. Admittedly, this line of research has stumbled on notable hurdles that might have limited its popularity \cite{randono:gauge}. Hence, adopting a novel outlook might be needed to try and overcome such issues. The central insight, which we build upon in this work, is that the mechanism allowing to recover a metric theory of gravity from the YM formalism is that of spontaneous symmetry breaking (SSB) \cite{hehl:metric,adler:einstein,bekenstein:gravitation}.\footnote{More recent studies aiming at connecting gravity and SSB in one way or another include {\it Krasnov}'s attempt to embed gravity into the mechanism of SSB of the Standard Model of particles \cite{krasnov:spontaneous}, {\it Chamseddine} \& {\it Mukhanov}'s formal unification of gauge and gravitational interactions via the gauge group $SO(1,13)$ \cite{chamseddine:unification} and {\it Westman} \& {\it Zlosnik}'s Cartan-geometric formulation of gravity as a gauge theory \cite{westman:cartan,Westman:2013mf,Westman:2014yca}. See also Refs.\ \cite{Zlosnik:2018qvg,Koivisto:2019ejt,Koivisto:2023epd,Koivisto:2022uvd,Koivisto:2024asr} for other attempts in model building and applications to cosmology and black holes.}\\

In this article, we present a pre-geometric gauge theory {\it à la} Yang--Mills out of which classical gravity, in the form of the Einstein--Cartan theory, dynamically emerges  due to a mechanism of SSB, while still attaining all the tenets upon which GR is based, including the principle of general covariance. By ``pre-geometric'' it is meant that no spacetime metric is introduced in the unbroken phase, foregoing the possibility of defining, for example, the notions of length and angle before the SSB has happened; on the other hand, in the spontaneously broken phase one can naturally identify all the geometric quantities that characterise the classical picture of spacetime, from the spacetime metric to the spin connection, from the Riemann curvature tensor to the EH action, thus justifying the expression ``emergent gravity'' that we will use to refer to the theory. In this sense, the emergence of the curvature of spacetime, as well as the dynamical laws that govern its evolution, is claimed to be a substantive property of the Universe and not a mere effective or analogical description resulting from the coarse-graining of some underlying degrees of freedom.\footnote{This is why the framework under consideration should not be confused with {\it Verlinde}'s theory of entropic gravity \cite{verlinde:gravity}, which is also referred to as emergent gravity, but is based on unrelated ideas inspired by the AdS/CFT correspondence. Rather, more relevant to the ideas presented in this article are the Refs.\ \cite{maitiniyazi:irreversible,maitiniyazi:generation,akama:pregeometry,wetterich:pregeometry,wetterich:spinors,diakonov:lattice,obukhov:diakonov,obukhov:dirac}.} Hence, the general-relativistic interpretation of gravity as the inertial effect of the curvature of spacetime on matter distributions is retained. Nonetheless, this simply ceases to make sense in the high-energy limit of the unbroken phase where spacetime is devoid of a metric structure.\\

Following the Refs.\ \cite{macdowell:unified,westman:cartan}, the fundamental gauge group of the four-dimensional spacetime manifold is taken to be either $SO(1,4)$ or $SO(3,2)$, with the mechanism of SSB reducing it to the stabiliser group $SO(1,3)$. This is suggestive of the origin of the Einstein equivalence principle (EEP) as due to the residual symmetry of the fundamental gauge principle of the unbroken phase. As a result of the SSB, the components of the pre-geometric YM gauge field decompose into two classes: those corresponding to the unbroken generators of the gauge group can be identified with the components of the spin connection, while those corresponding to the broken generators can be identified with the tetrads of the Palatini formalism. The components of the gauge connection in the unbroken phase are equal in number to those of the spin connection and the tetrads of the spontaneously broken phase. This thereby justifies the choice of the initial gauge group as the minimal extension of the Lorentz group that can unify the spin connection and the tetrads as the connection of a single `higher symmetry' gauge group. These identifications then allow to define a dictionary for emergent gravity, in order to `translate' from the pre-geometric fields to the relevant effective geometric quantities. Other than recovering the EH action with the cosmological constant term and showing that diffeomorphism invariance and the EEP can be emergent rather than fundamental, our construction implements two key improvements with respect to the previous literature. Compared to the works \cite{macdowell:unified} and \cite{westman:cartan}, the mechanism of SSB is made dynamical by the dynamics of a Higgs-like field, hence getting rid of {\it ad hoc} Lagrange multipliers or special gauge conditions. Second, our formalism is fully generally covariant and so is unconstrained by requirements such as the unimodular condition (differently from \cite{wilczek:gauge}). In view of these premises, we will argue that the formal structure of the gravitational sector of the theory is tightly constrained by fundamental symmetries and physical principles.\\

In this framework, the two mass scales that characterise GR, i.e.\ the Planck mass and the cosmological constant, are emergent, as no meaningful definition thereof is available in the unbroken phase. This is consistent with the idea that, before the SSB, \emph{any} quantity related to classical gravity is assumed not to exist. The sign of the cosmological constant is correctly obtained depending on whether the fundamental gauge group is the de Sitter or the anti-de Sitter group. Furthermore, our discussion of the matter couplings will clarify in which sense the validity of the EEP is possible only at a classical level, while being subject to tiny violations due to quantum effects. The phase transition from the unbroken to the spontaneously broken phase realised at a certain critical temperature by the Higgs-like field has potentially important consequences for cosmology, as it generates a mass for the quanta of some of the pre-geometric fields in the early Universe. Such critical temperature is expected to be that of the Planck epoch, depending on a free parameter. Intriguingly, some of these Beyond Standard Model particles resulting from the Higgs mechanism for gravity are predicted to be Planckian tachyons. Last but not least, a case is made for the potential power-counting renormalisability of the theory in the absence of matter, in what could prove to be a new approach to find a long sought-after UV completion of Quantum Gravity.\\

We will employ natural units throughout the paper, which is organised as follows. After outlining the metric and tetrad formalisms for gravity in Sec.\ \ref{sec:2}, we describe the spontaneously broken phase of the theory and discuss all emergent geometric quantities in Sec.\ \ref{sec:3}. The dictionary for emergent gravity is written up in Sec.\ \ref{sec:4} and then utilised to delineate the details respectively of the mechanism of SSB in Sec.\ \ref{sec:5}, the matter couplings in Sec.\ \ref{sec:6} and the kinematics of pre-geometric fields in Sec.\ \ref{sec:7}. The details of the Higgs mechanism for gravity are then elucidated in Sec.\ \ref{sec:8}. Following a brief reflection in Sec.\ \ref{sec:9} on the interpretation of the theory and some cosmological consequences thereof, in Sec.\ \ref{sec:10} we present a simple extension of the theory based on the gauge symmetry of the group $SO(2,4)$ in order to include the conformal symmetry. Finally, in Secs.\ \ref{sec:11} and \ref{sec:12} we summarise and discuss the main results, mentioning some possible directions for future developments. The appendices \ref{sec:appendix_A} and \ref{sec:appendix_B} contain the explicit computations needed to obtain the main results of the SSB and the Higgs mechanism respectively, while the App.\ \ref{sec:appendix_C} contains those needed for the study of the conformal symmetry.

\section{Einstein gravity}\label{sec:2}
\subsection{The metric formalism}
The EH Lagrangian density in the metric formalism is expressed by the relation
\begin{equation}
    \mathcal{L}_\textup{EH}=\frac{M_\textup{P}^2}{2}\sqrt{-g}R,
\end{equation}
where $M_\textup{P}$ is the reduced Planck mass and the Ricci scalar is obtained from two contractions of the Riemann tensor as
\begin{equation}
    R\equiv g^{\rho\mu}g^{\sigma\nu}R_{\rho\sigma\mu\nu}=R_{\mu\nu}^{\mu\nu},
\end{equation}
where $R_{\mu\nu}^{\mu\nu}\equiv R^{\mu\nu}_{\phantom{\mu\nu}\mu\nu}$, with the Riemann tensor defined, in GR, as the curvature of the Levi-Civita connection of the tangent bundle of a Lorentzian manifold.\\

The tangent bundle of a four-dimensional Lorentzian manifold is the disjoint union of all its tangent spaces, with each of them being isomorphic to a Minkowski spacetime, in agreement with the EEP of GR. Thus the corresponding endomorphism bundle is the Lie algebra of the Lorentz group $SO(1,3)$. The most general metric connection that can be defined on such tangent bundle is a spin (or Lorentz) connection, which can always be decomposed into the sum of the Levi-Civita connection, i.e.\ the unique symmetric connection which is compatible with the metric, and a contortion tensor. A spin connection can thus be regarded as the gauge potential of a local $SO(1,3)$ gauge symmetry, hinting at the possibility of realising the EEP as a gauge symmetry. Physically, this interpretation of the EEP would suggest that classical spacetime itself, as a dynamical system, has an internal, gauge symmetry such that the effects of the curvature of spacetime can be \emph{locally} neglected. This allows to reduce physical laws to those valid in the Minkowski spacetime, which are indeed covariant under the transformations of the Lorentz group $SO(1,3)$.\\

The other foundational principle of GR, the principle of general covariance, refers to the invariance of the form of physical laws under coordinate transformations, following the rules of Ricci calculus. These include the transformation of a given metric tensor, which allows to raise, lower and contract the spacetime indices of tensors. But if no spacetime metric is assumed to exist, then the same principle must be interpreted in a broader sense. The main hurdle is then that, without a metric tensor, the action principle cannot be formulated in the standard way, thus preventing the derivation of the fundamental laws of GR and of the Standard Model of particles from such a well-established paradigm.

\subsection{The tetrad formalism}
The EH Lagrangian density in the tetrad formalism is expressed as
\begin{equation}
    \mathcal{L}_\textup{EH}=\frac{M_\textup{P}^2}{2}ee_a^\mu e_b^\nu R_{\mu\nu}^{ab},
\end{equation}
with the Riemann tensor defined, more generally, as the curvature of the spin connection $\omega_\mu^{ab}$:
\begin{equation}
    R_{\mu\nu}^{ab}=\partial_\mu\omega_\nu^{ab}-\partial_\nu\omega_\mu^{ab}+\omega_{c\mu}^a\omega_\nu^{cb}-\omega_{c\nu}^a\omega_\mu^{cb}=2(\partial_{[\mu}\omega_{\nu]}^{ab}+\omega_{c[\mu}^a\omega_{\nu]}^{cb}),
\end{equation}
where $\omega^a_{c\mu}\equiv\omega^a_{\phantom{c}c\mu}$ and the contraction of the internal indices involves the structure constants of $SO(1,3)$. A spin connection is indeed a one-form assuming values in the Lie algebra of $SO(1,3)$, and the corresponding Riemann tensor is a two-form assuming values in the Lie algebra of $SO(1,3)$.\footnote{See the Ref.\ \cite{luongo:repulsive} for applications related to the irreducible representations of the Riemann tensor.} Note that the existence of an inverse of the tetrads is essential for expressing the EH Lagrangian density in this form, otherwise there would exist no way to contract the spacetime (Greek) indices of the curvature two-form with its internal-space (Latin) indices.\\

In constructing the EH action in the tetrad formalism, the spin connection can be made to appear only algebraically after an integration by parts (because the Riemann tensor is linear in the first derivatives of the spin connection), implying that it is possible to eliminate the spin connection from the gravitational action in favour of the tetrads or, ultimately, a metric. This situation is in stark contrast with what happens for conventional YM theories, where instead gauge potentials play a fundamental role. In addition to that, on account of the soldered character of the tangent bundle expressed by the condition
\begin{equation}\label{eq:soldered}
    \eta_{ab}e_\mu^a e_\nu^b=g_{\mu\nu},
\end{equation}
locally the tetrads add no gauge-invariant information to that contained in the metric. Thus the counting of the degrees of freedom of the gravitational field is the same in both the metric and tetrad formalism.\footnote{In general, for $n\in N$ number of spacetime dimensions, the gravitational degrees of freedom are found to be: in the metric formulation, $n(n+1)/2$ for the symmetric tensor $g_{\mu\nu}$;  in the tetrad formulation, $n^2-n(n-1)/2=n(n+1)/2$ for the $n^2$ tetrads $e_\mu^a$ upon taking into account the $n(n-1)/2$-dimensional Lie algebra of the $SO(1,n-1)$ gauge symmetry.} In any case, in what follows we will make use of the Palatini formalism, treating the tetrads and the spin connection as independent variables.

\section{Gravity in the spontaneously broken phase}\label{sec:3}
Let us consider a gauge field theory in a pre-geometric four-dimensional spacetime, the gauge group of which is either the de Sitter group $SO(1,4)$ or the anti-de Sitter group $SO(3,2)$, with the associated gauge potentials denoted as $A_\mu^{AB}$ and the corresponding gauge field strengths denoted as $F_{\mu\nu}^{AB}$. The gauge potentials are antisymmetric in the Latin indices and the gauge field strengths are antisymmetric in both the Latin and Greek indices. The goal of this construction is to obtain dynamically a metric structure {\it à la} Riemann and Einstein -- hence recovering both $\mathcal{L}_\textup{EH}$ and the EEP -- without introducing any spacetime metric nor tetrads, while complying with the principle of general covariance. Only the internal-space metric, which is a generalisation of the Minkowski metric $\eta$, is assumed to exist for each tangent space to the spacetime manifold, with signature respectively $(-,+,+,+,+)$ or $(+,+,+,-,-)$. To achieve this, one can resort to the Higgs mechanism by means of a spacetime scalar, internal-space vector field $\phi^A$, as the SSB of its ground state can reduce the gauge symmetry from $SO(1,4)$ or $SO(3,2)$ to $SO(1,3)$ \cite{wilczek:gauge}. As a result, the curvature of spacetime and its dynamics can be seen to be emergent in the spontaneously broken phase, effectively arising from the interactions of the pre-geometric fields $A_\mu^{AB}$ and $\phi^A$ below a certain energy scale or critical temperature.\\

Compliance with the principle of general covariance requires Lagrangian densities to be scalar densities of weight $-1$ \cite{carroll:spacetime}. Since $A_\mu^{AB}$ and $F_{\mu\nu}^{AB}$ are covariant tensors in their spacetime indices, one needs objects that are contravariant in such indices to contract them and form scalar densities. In the absence of an inverse metric tensor, the only four-dimensional contravariant object that is intrinsically defined on the spacetime manifold is the constant Levi-Civita symbol $\epsilon^{\mu\nu\rho\sigma}$, which is a tensor density of weight $-1$. By `intrinsically' it is meant that no additional information other than the differential structure of the spacetime manifold is required -- in particular, no metric structure is needed. Moreover, let us observe that the density character of tensor densities themselves does not require a metric to be defined, but only the Jacobian determinant of a generic diffeomorphism of coordinates. Therefore, assuming general covariance, the contravariant Levi-Civita symbol can be used to the first power only in forming Lagrangian densities (since it already has the appropriate weight), to contract exactly four covariant spacetime indices.\\

Making use of the Levi-Civita symbol, one can construct two different gravitational Lagrangian densities for the unbroken phase. The Lagrangian density introduced by {\it MacDowell} \& {\it Mansouri} \cite{macdowell:unified} is expressed by 
\begin{equation}
    \mathcal{L}_\textup{MM}=k_\textup{MM}\epsilon_{ABCDE}\epsilon^{\mu\nu\rho\sigma}F_{\mu\nu}^{AB}F_{\rho\sigma}^{CD}\phi^E,
\end{equation}
while that of {\it Wilczek} \cite{wilczek:gauge} is written in the form
\begin{equation}
    \mathcal{L}_\textup{W}=k_\textup{W}\epsilon_{ABCDE}\epsilon^{\mu\nu\rho\sigma}F_{\mu\nu}^{AB}\nabla_\rho\phi^C\nabla_\sigma\phi^D\phi^E,
\end{equation}
where $\nabla_\mu$ denotes the internal-space gauge covariant derivative, which acts on internal-space vectors as
\begin{equation}\label{eq:cov_dev}
    \nabla_\mu\phi^A=\partial_\mu\phi^A+A_{B\mu}^A\phi^B=(\delta^A_B\partial_\mu+A^A_{B\mu})\phi^B,
\end{equation}
with $A_{B\mu}^A\equiv\eta_{BC}A_\mu^{AC}$. Uppercase Latin indices run from $1$ to $5$ and Greek indices run from $1$ to $4$. Observe that the mass dimensions of the two coupling constants introduced above are respectively $[k_\textup{MM}]=[\phi]^{-1}$ and $[k_\textup{W}]=[\phi]^{-3}$, with the mass dimension of the field $\phi^A$ being unspecified {\it a priori}.\\

Before discussing how the field $\phi^A$ can give rise dynamically to a mechanism of SSB, in this section we will analyse what happens when such mechanism has already taken place and the gauge symmetry has been broken from $SO(1,4)$ or $SO(3,2)$ to $SO(1,3)$. In this way, we will first appreciate what the interest is in studying these theories from the point of view of classical physics, after which we will focus on setting up the exploration of the uncharted pre-geometric territory.

\subsection{Gravity {\it à la} MacDowell and Mansouri}
The mechanism of SSB will result in the emergence of a preferential direction in the internal space, which is singled out by the frozen value $\phi^A=v\delta^A_5$ for the field $\phi^A$, with $v$ a nonzero constant. Such preferential direction enables to distinguish between two classes of gauge potentials: for each value of the covariant spacetime index, there are four gauge potentials $A_\mu^{A5}\equiv A_\mu^{a5}$, and then six gauge potentials $A_\mu^{AB}\equiv A_\mu^{ab}$ (with neither $A$ nor $B$ equal to $5$) -- lowercase Latin indices running from $1$ to $4$ have been introduced to describe the spontaneously broken phase. First, let us carry out the computation of the SSB for the Lagrangian density $\mathcal{L}_\textup{MM}$. Thereafter essentially the same steps will allow to reproduce this computation effortlessly for $\mathcal{L}_\textup{W}$ as well. Expanding more explicitly the expression of $\mathcal{L}_\textup{MM}$ after the SSB, the $A_\mu^{a5}$ gauge potentials can be seen to appear only in spacetime commutator terms, namely
\begin{equation}
    \begin{split}
    \mathcal{L}_\textup{MM}&\xrightarrow{SSB}k_\textup{MM}v\epsilon_{ABCDE}\epsilon^{\mu\nu\rho\sigma}F_{\mu\nu}^{AB}F_{\rho\sigma}^{CD}\delta^E_5\\
    &=4k_\textup{MM}v\epsilon_{abcd}\epsilon^{\mu\nu\rho\sigma}(\partial_{[\mu}A_{\nu]}^{ab}+A_{E[\mu}^aA_{\nu]}^{Eb})(\partial_{[\rho}A_{\sigma]}^{cd}+A_{F[\rho}^cA_{\sigma]}^{Fd})\\
    &=4k_\textup{MM}v\epsilon_{abcd}\epsilon^{\mu\nu\rho\sigma}(\partial_{[\mu}A_{\nu]}^{ab}+A_{e[\mu}^aA_{\nu]}^{eb}\mp A_{[\mu}^{a5}A_{\nu]}^{b5})(\partial_{[\rho}A_{\sigma]}^{cd}+A_{f[\rho}^cA_{\sigma]}^{fd}\mp A_{[\rho}^{c5}A_{\sigma]}^{d5}).
    \end{split}
\end{equation}
In the last equality it was introduced the convention of denoting with the first sign the case of the $SO(1,4)$ gauge group and with the second sign the case of the $SO(3,2)$ gauge group.\\

Upon making the identifications
\begin{equation}\label{eq:identifications}
    e_\mu^a\equiv m^{-1}A_\mu^{a5},\qquad\omega_\mu^{ab}\equiv A_\mu^{ab},
\end{equation}
where the mass parameter $m$ has to be introduced for dimensional reasons, the Lagrangian density can be recast as a sum of three terms:
\begin{equation}\label{eq:SSB_MM}
    \begin{split}
    \mathcal{L}_\textup{MM}&\xrightarrow{SSB}k_\textup{MM}v\epsilon_{abcd}\epsilon^{\mu\nu\rho\sigma}(R_{\mu\nu}^{ab}\mp2m^2e_{[\mu}^ae_{\nu]}^b)(R_{\rho\sigma}^{cd}\mp2m^2e_{[\rho}^ce_{\sigma]}^d)\\
    &=k_\textup{MM}v(\epsilon_{abcd}\epsilon^{\mu\nu\rho\sigma}R_{\mu\nu}^{ab}R_{\rho\sigma}^{cd}\mp4m^2\epsilon_{abcd}\epsilon^{\mu\nu\rho\sigma}e_\mu^ae_\nu^bR_{\rho\sigma}^{cd}\\
    &+4m^4\epsilon_{abcd}\epsilon^{\mu\nu\rho\sigma}e_\mu^ae_\nu^be_\rho^ce_\sigma^d).
    \end{split}
\end{equation}

The remaining computations for each of these terms are displayed in the App.\ \ref{sec:appendix_A}, and lead to the result that the spontaneously broken phase of $\mathcal{L}_\textup{MM}$ is described by a Lagrangian density, which is equivalent to the EH one plus a cosmological constant term and a Gauss--Bonnet term (whose definition is given in the Eq.\ \eqref{eq:gauss-bonnet}):
\begin{equation}
    \mathcal{L}_\textup{MM}\xrightarrow{SSB}\pm16k_\textup{MM}vm^2ee_a^\mu e_b^\nu R_{\mu\nu}^{ab}-96k_\textup{MM}vm^4e-4k_\textup{MM}ve\mathcal{G}.
\end{equation}
For consistency with observations, this result requires the reduced Planck mass to be identified as
\begin{equation}
    M_\textup{P}^2\equiv\pm32k_\textup{MM}vm^2.
\end{equation}
In this sense, the Planck scale is only emergent in this theory, as it depends on a specific multiplicative combination of the constants $k_\textup{MM}$, $v$ and $m$. From this follows that the cosmological constant is emergent too, as it can be identified as
\begin{equation}
    \Lambda\equiv\pm3m^2=\frac{3M_\textup{P}^2}{32k_\textup{MM}v}.
\end{equation}
The last expression for the cosmological constant shows that it is naturally suppressed by a ``see-saw'' mechanism: setting the value of $M_\textup{P}^2$ equal to the one determined by the experimental infrared measurements of the gravitational constant ($M_\textup{P}^2\sim10^{37}$\,GeV\textsuperscript{2}) and assuming (even though this is not necessary) to have a coupling constant $k_\textup{MM}\sim\pm1\,[\phi]^{-1}$, a small value $\Lambda\sim10^{-84}$\,GeV\textsuperscript{2} can be obtained from a big value $v\sim10^{119}\,[\phi]$ -- plugging in the precise numbers, this is actually $v\approx10^{119}\,[\phi]$. This result provides a simple explanation for the smallness of the observed value of the cosmological constant, thereby avoiding any anthropic explanations \cite{weinberg:cosmological}. Thus, in this theory, the cosmological constant is determined by the mass parameter $m$ characterising the spontaneously broken phase. Given the observed value of the cosmological constant, this would imply $m\sim H_0$, that is a tiny energy scale of the order of the Hubble constant $H_0$.

\subsection{Gravity {\it à la} Wilczek}
The computation of the SSB for the Lagrangian density $\mathcal{L}_\textup{W}$ follows the same steps described for $\mathcal{L}_\textup{MM}$. It does require only one additional ingredient, which is the action of the covariant derivative on internal-space vectors, once the value $\phi^A=v\delta^A_5$ is fixed. From the Eq.\ \eqref{eq:cov_dev} one finds that
\begin{equation}\label{eq:cov-der}
    \nabla_\mu\phi^A\xrightarrow{SSB}v\nabla_\mu\delta^A_5=v(\cancel{\partial_\mu\delta^A_5}+A_{B\mu}^A\delta^B_5)=vA_{5\mu}^A=\pm vA_\mu^{a5}.
\end{equation}
Therefore, using the identifications \eqref{eq:identifications} and skipping the details of the computation, which are again displayed in the App.\ \ref{sec:appendix_A}, one obtains
\begin{equation}
    \mathcal{L}_\textup{W}\xrightarrow{SSB}-4k_\textup{W}v^3m^2ee^\mu_ae^\nu_bR_{\mu\nu}^{ab}\pm48k_\textup{W}v^3m^4e.
\end{equation}
In this theory, which corresponds to the EH Lagrangian density plus a cosmological constant term, the reduced Planck mass and the cosmological constant can be identified respectively as
\begin{equation}
    M_\textup{P}^2\equiv-8k_\textup{W}v^3m^2
\end{equation}
and
\begin{equation}
    \Lambda\equiv\pm6m^2=\mp\frac{3M_\textup{P}^2}{4k_\textup{W}v^3}.
\end{equation}
Note that this theory does not yield any Gauss--Bonnet term. Again, assuming to have a coupling constant $k_\textup{W}\sim-1\,[\phi]^{-3}$, a small value $\Lambda\sim10^{-84}$\,GeV\textsuperscript{2} can be obtained\footnote{It is conceivable that at the quantum level the coupling constants $k_\textup{MM}$ and $k_\textup{W}$ will run with the energy scale driven by the underlying renormalisation group flow. This would imply that the Planck mass $M_\textup{P}$ and the gravitational constant will run with the energy scale too. Nevertheless, we expect that such running will not affect the cosmological constant $\Lambda$ in both theories $\mathcal{L}_\textup{MM}$ and $\mathcal{L}_\textup{W}$.} from a big value $v\sim10^{40}\,[\phi]$.\\

In Sec.\ \ref{sec:11} we will provide an argument for discriminating between $\mathcal{L}_\textup{MM}$ and $\mathcal{L}_\textup{W}$ on the basis of observations and internal consistency; this result also hinges on the discussion presented in Subsec.\ \ref{subsec:3.5}, and could thus be generalised. Anyway, both theories have three free parameters, namely $m$, $v$ and $k_\textup{MM}$ or $k_\textup{W}$ respectively, which are constrained by the measured values of two constants, that is $M_\textup{P}^2$ and $\Lambda$. Therefore, only one of these three parameters is actually free in each theory, with both $M_\textup{P}^2$ and $\Lambda$ being independent of the mass dimension of the field $\phi^A$. In addition, it is interesting to observe that both theories under examination forbid the existence of the anti-de Sitter space as a solution of the classical field equations for gravity if the fundamental gauge group is $SO(1,4)$ -- since they predict $\Lambda>0$ -- and, vice versa, they forbid the existence of the de Sitter space in the case of $SO(3,2)$ -- since they predict $\Lambda<0$. In any case, the existence of a nonzero cosmological constant is an unavoidable result.\\

Observe that, as far as the values of $M_\textup{P}^2$ and $\Lambda$ are concerned, there exists a degeneracy between the possible values of $v$ and $k_\textup{MM}$ or $k_\textup{W}$: the bigger the former, the smaller the latter and vice versa. In what follows, we will assume for definiteness that $\lvert k_\textup{MM}\rvert\lesssim1\,[\phi]^{-1}$ and $\lvert k_\textup{W}\rvert\lesssim1\,[\phi]^{-3}$, a possibility inspired by perturbative approaches to field theories. Of course, there is no fundamental reason why the opposite possibility should not be correct, in which case non-perturbative approaches would have to be considered.

\subsection{Gravitational gauge potentials}\label{subsec:3.3}
According to the proposed identifications \eqref{eq:identifications} of the pre-geometric gauge fields, for each value of the covariant spacetime index, the $n(n-1)/2=10$ gauge potentials of the $SO(1,4)$ or $SO(3,2)$ gauge symmetry in $n=5$ internal-space dimensions decompose exactly into the $n=4$ tetrads and the $n(n-1)/2=6$ internal-space components of the spin connection associated with the $SO(1,3)$ gauge symmetry in $n=4$ internal-space dimensions. Therefore, to sum up, both theories $\mathcal{L}_\textup{MM}$ and $\mathcal{L}_\textup{W}$ exploit a mechanism of SSB to pass from the pre-geometric gauge fields $A_\mu^{AB}$ associated with the gauge symmetry $SO(1,4)$ or $SO(3,2)$ to exactly the geometric fields $e_\mu^a$ and $\omega_\mu^{ab}$ associated with the gauge symmetry $SO(1,3)$. This leads to the emergence of a classical metric structure on a four-dimensional spacetime manifold, which obeys the laws of the Einstein--Cartan theory. The non-metric, pre-geometric spacetime of the unbroken phase can \emph{locally} be thought of as having the symmetries of a flat spacetime with one more spatial or time dimension.\\

In the identifications \eqref{eq:identifications} lies indeed one of the most important conceptual merits of the present construction. In the Palatini formalism of the Einstein--Cartan theory, one of the two fundamental fields associated with the gravitational interaction is not a gauge potential of the Lorentz group: given a local Lorentz transformation represented by a matrix $\Lambda_b^a(x)$, in fact, the spin connection transforms inhomogeneously while the tetrads transform homogeneously, as
\begin{equation}
    \omega_\mu^{ab}\rightarrow\Lambda_c^a\omega_\mu^{cd}(\Lambda^{-1})_d^b+\eta^{cd}\Lambda_c^a\partial_\mu(\Lambda^{-1})_d^b,\qquad e_\mu^a\rightarrow\Lambda_b^ae_\mu^b.
\end{equation}
Therefore, the tetrads cannot play the role of gauge potentials in conventional attempts to implement the gauge principle into classical gravity, albeit at the same time they are what comprises the metric information content of the theory. This puzzling difference is key in understanding why gravity is so hard to be recast as a YM theory. The approach that we are using, instead, tackles this conundrum from a considerably different perspective. In passing from the $SO(1,4)$ or $SO(3,2)$ to the $SO(1,3)$ gauge symmetry for spacetime, the gauge potentials $A_\mu^{AB}$ of the unbroken phase reduce to the gauge potentials $A_\mu^{ab}$ of the spontaneously broken phase, which are identified with the spin connection, while the remaining fields $A_\mu^{a5}$ are identified with the tetrads which are not gauge potentials. This scenario is not only consistent, but also natural, because the fields $A_\mu^{a5}$ are gauge potentials only of the initial gauge symmetry but not of the residual one that is obtained via the SSB. After the phase transition from the pre-geometric to the metric spacetime, the fields $A_\mu^{ab}$ maintain their gauge character in terms of transformation laws, while the fields $A_\mu^{a5}$ lose it because they correspond to the broken generators of the initial symmetry group. Thus, the fields $A_\mu^{a5}$ \emph{must} be identified with gravitational degrees of freedom that do \emph{not} transform as gauge potentials of the Lorentz group. This justifies the introduction of the tetrads in the identifications \eqref{eq:identifications}.\\

In light of the above considerations, the choice of the initial gauge group as either $SO(1,4)$ or $SO(3,2)$ can be reinterpreted and corroborated as follows. If one tries to start from the Lorentz group $SO(1,3)$ and the EH action to implement the gauge principle into gravity, they will face the difficulty of not being able to treat the tetrads as gauge potentials. One way to surmount the problem of the stark difference between the tetrads and the spin connection is aiming to place them ``under the same umbrella'' by extending the gauge group. The natural choice is then either the de Sitter or the anti-de Sitter group because, as discussed at the beginning of this subsection, they are the two possible minimal extensions of the Lorentz group that allow to unify the spin connection and the tetrads as the components of a single gauge potential of a `higher symmetry' \cite{randono:gauge}. However, it must be kept in mind that this is achieved in a pre-geometric gauge field theory and not in a metric theory of gravity. Any bigger gauge group would yield, after the SSB, additional fields that would be hard to interpret in terms of gravitational degrees of freedom.\\

In a nutshell, this approach recovers classical gravity from the gauge principle (plus SSB) of a formulation {\it à la} Yang--Mills that is {\it a priori} unrelated to gravity, rather than implementing the gauge principle into the formalism of a metric theory of gravity. Moreover, the degrees of freedom of the pre-geometric gauge field correctly reproduce those of gravity in the spontaneously broken phase.

\subsection{The emergence of diffeomorphism invariance}
A metric structure is not the only classical feature of the world that emerges in this framework: diffeomorphism invariance and, in a sense, the principle of background independence are emergent too. At this point, it is important to make a distinction between three principles that are usually intended as interchangeable within the context of GR. In a theory of emergent gravity formulated on a pre-geometric spacetime, in fact, general covariance, background independence and diffeomorphism invariance are not equivalent principles. Before clarifying this subtle aspect, let us briefly recall the two possible types of realisations, or interpretations, of any given transformation on the spacetime manifold \cite{gaul:diffeomorphism}. \textit{Passive} diffeomorphisms are coordinate transformations (or changes of `reference frame' in the physicists' parlance) that express the same physical quantity with different coordinates. \textit{Active} diffeomorphisms, instead, are actual transformations of a physical quantity into another while leaving the coordinate system unchanged. By way of illustration, a passive diffeomorphism can be pictured as viewing a cube from different points of view, while an active diffeomorphism would correspond to, say, rotating the cube while keeping fixed the observer's perspective. In the previous example, of course, the distinction between the two types of diffeomorphisms is purely conceptual and lexical, but not physical; the difference becomes physically relevant only for transformations of background fields, like the spacetime metric. In GR, an example of passive diffeomorphism is a change of coordinates in the Schwarzschild spacetime (Schwarzschild coordinates, Eddington--Finkelstein coordinates, Kruskal--Szekeres coordinates etc.), which does not affect the physical properties of such spacetime (like its symmetries, its matter source etc.). An example of active diffeomorphism in GR is instead any transformation of a physical quantity, e.g.\ a spacetime interval, due to the use of \emph{different} metrics on the same spacetime while keeping the coordinates unvaried; this could be, for instance, the spacetime distance between two points in the Minkowski and the Friedmann--Lema\^{i}tre--Robertson--Walker metrics both in Cartesian coordinates $(t,x,y,z)$ or in the Minkowski and the Schwarzschild metrics both in spherical coordinates $(t,r,\theta,\varphi)$: an active diffeomorphism relating the points of the spacetime manifold will result, in general, in a different spacetime distance between any pair of points when computed with different metrics.\\

Any physical theory can be either invariant under passive diffeomorphisms or not and, independently of that, invariant under active diffeomorphisms or not. Invariance under passive diffeomorphisms is what we will refer to as \emph{general covariance} (sometimes also called general principle of relativity). This is merely a formal (and possibly even trivial) property of those physical laws that are valid for any choice of reference frame, but it is not a property of the Universe. For instance, the field equations of GR are usually formulated in a coordinate-free form, which is what is meant by saying that the Einstein equations are generally covariant, though they can also be specialised to some specific coordinate systems, in which case their manifest general covariance goes missing. On the other hand, invariance under active diffeomorphisms is what we will refer to as \emph{diffeomorphism invariance} (acknowledging its use within the Quantum Gravity community), a principle that can be understood in terms of symmetries since the dynamics of diffeomorphism invariant theories is independent of active diffeomorphisms that generate new fields, like new gauge fields or new background metrics. For example, in GR it is always possible to find a transformation relating the points of the spacetime manifold such that the physical distance between any pair of them is the same when computed with different metrics, with each of such metrics being a solution of the Einstein equations. Finally, especially when referring to geometric quantities, \emph{background independence} is the principle according to which the dynamics of all (matter or geometric) fields is independent of a background metric or, in other words, the metric tensor itself is dynamical rather than fixed -- and this is a property of the Universe regardless of the form of the equations employed to describe it. In the case of GR, the theory is background independent because it describes the spacetime metric as a dynamical field, the evolution of which is subject to the Einstein equations. At the same time, it is also diffeomorphism invariant, because if a metric is a solution of the field equations, then any other metric that can be related to it via an active diffeomorphism will still be a solution of the same equations. In GR, diffeomorphism invariance is ensured by general covariance because the spacetime metric itself is not a fixed background; this is exemplified, for instance, by the fact that the Jacobian determinant $J$ of a generic change of coordinates is related to the metric determinant by the relation $J=1/\sqrt{-g}$, which allows to recast the generally covariant volume element of spacetime, and thus the action principle, in a coordinate-free form. This formal result follows from the physical property of the invariance of spacetime intervals under both passive and active diffeomorphisms.\\

When it comes to the theories of emergent gravity under consideration, while it is obvious that the Lagrangian densities derived in the spontaneously broken phase are diffeomorphism invariant, as highlighted in particular by the presence of the factor $e=\sqrt{-g}$, such principle cannot even be expressed in pre-geometric terms, as a spacetime metric is not defined in the unbroken phase. In any case, both Lagrangian densities $\mathcal{L}_\textup{MM}$ and $\mathcal{L}_\textup{W}$ are manifestly generally covariant, because the contravariant spacetime Levi-Civita symbol $\epsilon^{\mu\nu\rho\sigma}$ provides the correct density character as the tetrad determinant $e=\sqrt{-g}$ does under general coordinate transformations. Therefore, in these formulations general covariance by no means implies the validity of diffeomorphism invariance, with the latter emerging in such generally covariant formalism only \emph{after} the SSB of a fundamental gauge symmetry of spacetime. As for the principle of background independence, the case is more nuanced. While background independence in the \emph{metric} sense is of course emergent, there is a pre-geometric or \emph{gauge} sense in which it is already present before the SSB: the dynamics of all fields in the unbroken phase, in fact, includes that of the gauge connection at each tangent space of the spacetime manifold, since the pre-geometric gauge fields $A_\mu^{AB}$ are themselves dynamical and do not act as a fixed background.

\subsection{On the uniqueness of the construction}\label{subsec:3.5}
The field theory presented in this section is formulated on a four-dimensional spacetime manifold. This comes equipped only with the Levi-Civita symbol $\epsilon^{\mu\nu\rho\sigma}$ (intrinsically defined on it), while its tangent spaces admit both the Levi-Civita symbol $\epsilon_{ABCDE}$ and the Minkowski metric $\eta_{AB}$. Therefore, the most glaring difference with respect to the spacetime of any metric theory of gravity, like GR, is the absence of a dynamical background metric. That said, the form of the possible Lagrangian densities ($\mathcal{L}_\textup{MM}$ or $\mathcal{L}_\textup{W}$) and the gauge group of the theory ($SO(1,4)$ or $SO(3,2)$) for the SSB leading to the emergence of gravity are far from being serendipitous. Quite the contrary, we will now proceed to show that both are \emph{unique} under a set of reasonable assumptions. More precisely, a mechanism of SSB of a fundamental gauge symmetry of spacetime for the recovery of classical gravity can be formally realised by only two Lagrangian densities, $\mathcal{L}_\textup{MM}$ or $\mathcal{L}_\textup{W}$, and only two gauge groups, $SO(1,4)$ or $SO(3,2)$, if all of the following hypotheses are satisfied:
\begin{enumerate}[I)]
    \item spacetime is four-dimensional;
    \item the principle of general covariance holds;
    \item the EH Lagrangian density $\mathcal{L}_\textup{EH}$ and the EEP are recovered in the spontaneously broken phase;
    \item the pre-geometric field content consists only of a gauge field $A_\mu^{AB}$ and a Higgs-like field $\phi^A$.
\end{enumerate}

The assumptions I and II imply that, in the absence of a spacetime metric, any possible Lagrangian density must contain one and only one contravariant Levi-Civita symbol $\epsilon^{\mu\nu\rho\sigma}$, as discussed at the beginning of this section. Additional spacetime Levi-Civita symbols would break general covariance, while increasing the number of spacetime dimensions would require a compactification of the extra dimensions to avoid inconsistencies with inverse-square laws. Given that the gauge principle concerns only internal symmetries, a mechanism of SSB can only help in reducing the dimensions of the internal space but not those of spacetime, since that would demand a spacetime \emph{vector} Higgs-like field ($\phi_\mu$) which would be at odds with the principle of local Lorentz invariance. The four contravariant spacetime indices of $\epsilon^{\mu\nu\rho\sigma}$ must be contracted with as many covariant spacetime indices coming from the pre-geometric field content, which is assumed to satisfy the hypothesis IV in order to implement the highest possible level of simplicity.\\

Covariant derivatives of $A_\mu^{AB}$ or $\nabla_\mu\phi^A$ can be excluded because they would be at variance with the assumption III and terms like $\propto\epsilon_{ABCDE}\epsilon^{\mu\nu\rho\sigma}F_{\mu\nu}^{AB}\nabla_{[\rho}\nabla_{\sigma]}\phi^C\phi^D\phi^E$, for example, are identically vanishing because of the antisymmetric character of the Levi-Civita symbol. That said, there exist only three possible combinations of the pre-geometric fields that can provide four covariant spacetime indices, namely two $F_{\mu\nu}^{AB}$'s, one $F_{\mu\nu}^{AB}$ and two $\nabla_\mu\phi^A$'s or four $\nabla_\mu\phi^A$'s: the first two combinations allow for the recovery of $\mathcal{L}_\textup{EH}$ as seen, for example, with $\mathcal{L}_\textup{MM}$ and $\mathcal{L}_\textup{W}$ respectively, while the third one can recover a cosmological constant term only, say $\mathcal{L}_J=k_J\lvert J\rvert$ with $k_J$ constant. In principle, these three terms can be considered simultaneously in a sum, as that would represent the most general Lagrangian density that can be constructed for an emerging gravitational interaction; at the same time, however, that would also mean that the two energy scales of classical gravity ($M_\textup{P}^2$ and $\Lambda$) are to be determined by combinations of the five parameters of the theory ($k_\textup{MM}$, $k_\textup{W}$, $k_J$, $v$ and $m$), resulting in a considerably reduced predictive power compared to the case discussed in this section with only three parameters. For this reason, in this work we will deliberately restrict our attention to the analysis of such terms \emph{individually}, thus neglecting $\mathcal{L}_J$ because it does not comply with the hypothesis III on its own. Finally, the only element of a possible Lagrangian density that remains to be discussed is related to the internal-space Levi-Civita symbol. The hypotheses III and IV imply that its rank must be five. In fact, having one Higgs-like field to break more than one internal-space dimension -- so as to still recover the $SO(1,3)$ gauge symmetry after the SSB -- means that there would be more than four broken generators and thus additional fields other than the tetrads -- this is in contrast with the argument presented in Subsec.\ \ref{subsec:3.3} on the unification of the spin connection and the tetrads into a single pre-geometric gauge field. This condition translates directly into a constraint on the choice of the gauge group of the theory, as the indefinite special orthogonal group $SO(p,q)$ (with $p,q\geq1$) admits a five-dimensional fundamental representation \emph{only} in two cases: the de Sitter group with $p=1,q=4$ and the anti-de Sitter group with $p=3,q=2$.\\

The number of internal-space Levi-Civita symbols to be used, instead, is arbitrary, but such choice either leads to a contradiction of the assumption III or to Lagrangian densities that are physically equivalent to $\mathcal{L}_\textup{MM}$ or $\mathcal{L}_\textup{W}$. To see this, let us discuss this choice in steps of increasing complexity. If only one internal-space Levi-Civita symbol is used, then only two Lagrangian densities are possible, namely $\mathcal{L}_\textup{MM}$ and $\mathcal{L}_\textup{W}$. Alternatively, multiple internal-space Levi-Civita symbols can be introduced by contracting some of their indices with internal-space Minkowski metrics. Thereby all four internal-space indices coming from the pre-geometric field content are still contracted. While no more than one field $\phi^A$ is contracted with each internal-space Levi-Civita symbol -- this ensures that the indices of each internal-space Minkowski metric are contracted with indices of two different internal-space Levi-Civita symbols. If two internal-space Levi-Civita symbols are used, then only one Lagrangian density is possible, which is
\begin{equation*}
    \propto\eta^{CH}\eta^{DI}\epsilon_{ABCDE}\epsilon_{FGHIJ}\epsilon^{\mu\nu\rho\sigma}F_{\mu\nu}^{AB}\nabla_\rho\phi^F\nabla_\sigma\phi^G\phi^E\phi^J.
\end{equation*}
Nonetheless, this latter contradicts the hypothesis III, since it does not allow to recover $\mathcal{L}_\textup{EH}$ after the SSB -- in the spontaneously broken phase, in fact, this is $\propto\eta^{eg}\eta^{fh}\epsilon_{abef}\epsilon_{cdgh}\epsilon^{\mu\nu\rho\sigma}e_\mu^ae_\nu^bR_{\rho\sigma}^{cd}$ instead of $\propto\epsilon_{abcd}\epsilon^{\mu\nu\rho\sigma}e_\mu^ae_\nu^bR_{\rho\sigma}^{cd}$. If three or more internal-space Levi-Civita symbols are used, then it is possible to contract them in such a way that $\mathcal{L}_\textup{EH}$ is recovered after the SSB, but such Lagrangian densities are physically indistinguishable from $\mathcal{L}_\textup{MM}$ or $\mathcal{L}_\textup{W}$ (except for the number of factors of $\phi^A$ employed, which is never bigger than two in any case). Basic examples are:
\begin{equation*}
    \propto\eta^{EJ}\eta^{FK}\eta^{GL}\eta^{HM}\eta^{IN}\epsilon_{ABCDE}\epsilon_{FGHIJ}\epsilon_{KLMNO}\epsilon^{\mu\nu\rho\sigma}F_{\mu\nu}^{AB}\nabla_\rho\phi^C\nabla_\sigma\phi^D\phi^O,
\end{equation*}
with three internal-space Levi-Civita symbols (and one factor of $\phi^A$), and
\begin{equation*}
    \propto\eta^{EJ}\eta^{FK}\eta^{GL}\eta^{HP}\eta^{IQ}\eta^{MR}\eta^{NS}\epsilon_{ABCDE}\epsilon_{FGHIJ}\epsilon_{KLMNO}\epsilon_{PQRST}\epsilon^{\mu\nu\rho\sigma}F_{\mu\nu}^{AB}\nabla_\rho\phi^C\nabla_\sigma\phi^D\phi^O\phi^T,
\end{equation*}
with four internal-space Levi-Civita symbols (and two factors of $\phi^A$).\\

In a way, the argument presented in this subsection parallels the approach of Lovelock's theorem \cite{lovelock:einstein}, which states the conditions for the uniqueness of the Lagrangian density for a metric theory of gravity. Specifically, Lovelock's theorem concerns the \emph{spacetime} features of geometric tensor fields, as it hinges crucially on the assumptions of the dimensionality of spacetime, locality in spacetime and conservation of energy and momentum in spacetime. The argument of this subsection aims at constraining the conditions for the uniqueness of a Lagrangian density for a theory of emergent gravity, via a mechanism of SSB. Thus it is mainly concerned with the \emph{internal-space} features of pre-geometric tensor fields, in particular the structure of the gauge group at each tangent space and the pattern of symmetry breaking. Here locality was not explicitly assumed as a hypothesis given that, to recover the local Lagrangian density $\mathcal{L}_\textup{EH}$, it is intuitive to start from a Lagrangian density for the unbroken phase, which is local too. The key difference with respect to Lovelock's theorem is of course the SSB related to gravity: if the classical metric theory of gravity is conceived as being emergent rather than fundamental, then our argument provides a natural generalisation of Lovelock's one for the uniqueness of the dynamics in the pre-geometric phase. As a concluding remark, we observe that the only way to consider extensions of the fundamental gauge group in this framework is to break one or more of the above assumptions, especially the hypothesis IV.

\section{A dictionary for emergent gravity}\label{sec:4}
In order to show that a metric theory of gravity can emerge out of a non-metric field theory in a complete and consistent way, it is essential to derive in the same formalism not only an effective metric $g_{\mu\nu}$, but also an effective inverse of the metric $g^{\mu\nu}$, as well as the tetrad determinant $e=\sqrt{-g}$. The inverse metric is required to construct most of the kinetic and interaction terms of GR and the Standard Model of particles in the Lagrangian formalism. Instead, the metric determinant is needed to formulate an action principle with a background-independent volume element. Making use of the fundamental relation \eqref{eq:soldered}, the metric can be built out of quadratic combinations of tetrads. Analogously, the inverse metric can be built out of quadratic combinations of inverse tetrads. In this section we expand upon the exposition of the Ref.\ \cite{wilczek:gauge}.\\

The definition of an emergent tetrad has been given in the identifications \eqref{eq:identifications}, which in turn allows to construct an effective metric as follows:
\begin{equation}
    m^{-2}\eta_{AB}A_\mu^{A5}A_\nu^{B5}\xrightarrow{SSB}\eta_{ab}e_\mu^ae_\nu^b\equiv g_{\mu\nu}.
\end{equation}
In light of the expression \eqref{eq:cov-der}, an effective metric can also be written as
\begin{equation}
    v^{-2}m^{-2}P_{\mu\nu}\xrightarrow{SSB}\eta_{ab}e_\mu^ae_\nu^b\equiv g_{\mu\nu},
\end{equation}
with
\begin{equation}\label{eq:aux_P}
    P_{\mu\nu}\equiv\eta_{AB}\nabla_\mu\phi^A\nabla_\nu\phi^B.
\end{equation}

To obtain a suitable expression for an inverse tetrad, one can first define the auxiliary field
\begin{equation}\label{eq:aux_w}
    w_A^\mu\equiv\pm\epsilon_{ABCDE}\epsilon^{\mu\nu\rho\sigma}\nabla_\nu\phi^B\nabla_\rho\phi^C\nabla_\sigma\phi^D\phi^E.
\end{equation}
In fact, after the SSB, this field becomes
\begin{equation}
    w_A^\mu\xrightarrow{SSB}v^4\epsilon_{ABCDE}\epsilon^{\mu\nu\rho\sigma}\delta_5^EA_\nu^{B5}A_\rho^{C5}A_\sigma^{D5}=v^4m^3\epsilon_{abcd}\epsilon^{\mu\nu\rho\sigma}e_\nu^be_\rho^ce_\sigma^d,
\end{equation}
which is effectively proportional to the inverse tetrad times the tetrad determinant, as it can be checked by considering that $e_a^\mu e_\lambda^a=\delta_\lambda^\mu$ (see also the Eq.\ \eqref{eq:id_cc}):
\begin{equation}\label{eq:w}
    w_a^\mu e_\lambda^a\xrightarrow{SSB}v^4m^3\epsilon_{abcd}\epsilon^{\mu\nu\rho\sigma}e_\lambda^ae_\nu^be_\rho^ce_\sigma^d=-6v^4m^3e\delta_\lambda^\mu.
\end{equation}
The latest result seems to suggest that the inverse tetrad and the tetrad determinant are not independent quantities. Thankfully, this is not the case, as an effective tetrad determinant can be constructed in an alternative and independent way via another auxiliary field, namely
\begin{equation}\label{eq:aux_J}
    J\equiv\epsilon_{ABCDE}\epsilon^{\mu\nu\rho\sigma}\nabla_\mu\phi^A\nabla_\nu\phi^B\nabla_\rho\phi^C\nabla_\sigma\phi^D\phi^E,
\end{equation}
which can be thought of as an effective Jacobian determinant. This is true because after the SSB one finds -- see, again, Eq.\ \eqref{eq:id_cc} -- that
\begin{equation}\label{eq:J}
    J\xrightarrow{SSB}v^5\epsilon_{ABCDE}\epsilon^{\mu\nu\rho\sigma}\delta_5^EA_\mu^{A5}A_\nu^{B5}A_\rho^{C5}A_\sigma^{D5}=v^5m^4\epsilon_{abcd}\epsilon^{\mu\nu\rho\sigma}e_\mu^ae_\nu^be_\rho^ce_\sigma^d=-24v^5m^4e.
\end{equation}
Finally, an effective inverse of the metric can be constructed as follows:
\begin{equation}\label{eq:inverse_metric}
    16v^2m^2J^{-2}\eta^{AB}w_A^\mu w_B^\nu\xrightarrow{SSB}\eta^{ab}e_a^\mu e_b^\nu\equiv g^{\mu\nu}.
\end{equation}

An inspection of the expressions \eqref{eq:w} and \eqref{eq:J} shows that, whenever it happens that $e=0$ at a given spacetime point, both auxiliary fields $w_A^\mu$ and $J$ will vanish. This in turn forbids to write down the expression \eqref{eq:inverse_metric} for the effective inverse of the metric. This situation is not different from what already happens in the Einstein--Cartan theory, where the EH action itself is proportional to $e$. Unfortunately, there is no straightforward way to dynamically enforce the non-degeneracy of the tetrads,\footnote{A possible but unsatisfactory way to circumvent this issue is to resort to the unimodular condition \cite{wilczek:gauge}.} so as to make sure that they are \emph{always} invertible. Nonetheless, this procedure is compatible with the principle of general covariance. However, assuming that the inverse of the effective tetrads exists is no further assumption than the one already implicitly made in the Einstein--Cartan theory for the inverse of the metric. So from this point of view \emph{both} the pre-geometric theory and the Einstein--Cartan theory will fail whenever the assumption of a non-singular metric fails and they cannot be discriminated on this aspect. This is also true because, in any case, coordinate singularities for which $e=0$ do not correspond to physical singularities since tetrads are not observable quantities.\\

To summarise, in this section we have built a comprehensive dictionary to translate from the language of the pre-geometric fields $A_\mu^{AB}$ and $\phi^A$ to that of the effective geometric fields $e_\mu^a$ and $g_{\mu\nu}$ after the SSB, which we will show again here for conciseness:
\begin{equation}\label{eq:dictionary}
    \begin{split}
    m^{-1}A_\mu^{a5}\,\text{or}\,\pm v^{-1}m^{-1}\nabla_\mu\phi^A&\xrightarrow{SSB}e_\mu^a,\\
    4vmJ^{-1}w_A^\mu&\xrightarrow{SSB}e_a^\mu,\\
    -\frac{1}{24}v^{-5}m^{-4}J&\xrightarrow{SSB}e=\sqrt{-g},\\
    m^{-2}\eta_{AB}A_\mu^{A5}A_\nu^{B5}\,\text{or}\,v^{-2}m^{-2}P_{\mu\nu}&\xrightarrow{SSB}\eta_{ab}e_\mu^ae_\nu^b\equiv g_{\mu\nu},\\
    16v^2m^2J^{-2}\eta^{AB}w_A^\mu w_B^\nu&\xrightarrow{SSB}\eta^{ab}e_a^\mu e_b^\nu\equiv g^{\mu\nu},
    \end{split}
\end{equation}
with the auxiliary fields $P_{\mu\nu}$, $w_A^\mu$ and $J$ defined in the Eqs.\ \eqref{eq:aux_P}, \eqref{eq:aux_w} and \eqref{eq:aux_J} respectively. It is straightforward to prove that all the effective geometric quantities are actual tensors and that, in the case of the effective metric and its inverse, they are also symmetric tensors. Remarkably, all effective geometric quantities of the unbroken phase can be constructed using only the field $\phi^A$, which is the one determining the mechanism of SSB by assuming the value $\phi^A=v\delta^A_5$, thus making the use of the gauge fields $A_\mu^{AB}$ redundant in this dictionary thanks to the formula \eqref{eq:cov-der}.

\section{Gravity from spontaneous symmetry breaking}\label{sec:5}
\subsection{Symmetry-breaking potential and unitary gauge}
The dynamical mechanism of SSB from the $SO(1,4)$ or $SO(3,2)$ gauge symmetry to the $SO(1,3)$ one via the field $\phi^A$, which leads to the emergence of a classical metric structure in spacetime, can be described by adding the following term to the Lagrangian density, which provides the simplest out of all possible symmetry-breaking potentials:
\begin{equation}\label{eq:potential}
    \mathcal{L}_\textup{SSB}=-k_\textup{SSB}v^{-4}\lvert J\rvert(\eta_{AB}\phi^A\phi^B\mp v^2)^2,
\end{equation}
with $k_\textup{SSB}$ a positive constant and $[k_\textup{SSB}]=[\phi]^{-5}$. This term is stationarised, and the potential $-\mathcal{L}_\textup{SSB}$ minimised, for instance when $\phi^A=v\delta_5^A$ (as $J\rightarrow\infty$, such value is effectively frozen in); any other possible solution for the vacuum expectation value (v.e.v.) of the field $\phi^A$ can be put in this form with a gauge transformation \cite{wilczek:gauge}. The factor $J$ ensures that $\mathcal{L}_\textup{SSB}$ is a scalar density, which guarantees that the principle of general covariance is obeyed. Notice, however, that $\lvert J\rvert$ can be disregarded from the Lagrangian density if the unimodular condition is applied, as {\it Wilczek} actually did \cite{wilczek:gauge}; in that case, the consistent dimensional choice for the coupling constant is $[k_\textup{SSB}]=[M]^4$ independently of the physical dimensions of $\phi^A$.\\

The quantisation of the field $\phi^A$ can be finally performed around its v.e.v. $\phi^A=v\delta_5^A$. In the unitary gauge, it reads $\phi^A=(v+\rho)\delta_5^A$. The unitary gauge is employed to suppress four Goldstone bosons from the definition of $\phi^A$. This expression for $\phi^A$ will be used to realise a Higgs mechanism for gravity in Sec.\ \ref{sec:8}.\\

It is important to observe that this mechanism of SSB is actually dynamical only if the field $\phi^A$ is itself dynamical, so that it can effectively ``roll down'' the potential given by $-\mathcal{L}_\textup{SSB}$, otherwise adding the term $\mathcal{L}_\textup{SSB}$ to the Lagrangian density is tantamount to adding a Lagrange multiplier to the theory -- as it could possibly be the case in the work \cite{wilczek:gauge}. The addition of a kinetic term for $\phi^A$ thus appears to be a necessity for the theory $\mathcal{L}_\textup{MM}$, otherwise the mechanism of SSB would be non-dynamical. On the other hand, within the case of $\mathcal{L}_\textup{W}$, $\phi^A$ is already dynamical and so in principle the term $\mathcal{L}_\textup{W}$ alone could already suffice for this purpose. In any case, the construction of an appropriate kinetic term for the field $\phi^A$ will be addressed in Sec.\ \ref{sec:7}, once a deeper understanding of the role and the meaning of emergence in these theories has been acquired.

\subsection{Classical and quantum effective geometric quantities}
The expressions for the effective geometric quantities given in the dictionary \eqref{eq:dictionary} are \emph{classical} in the sense that they ignore the fluctuations of the fields $A_\mu^{AB}$ and $\phi^A$ once these are quantised, leading to
\begin{equation}
    A_\mu^{ab}\equiv a_\mu^{ab}+\sigma_\mu^{ab}\xrightarrow{SSB}\omega_\mu^{ab}+\sigma_\mu^{ab},\qquad A_\mu^{a5}\equiv a_\mu^a+\sigma_\mu^a\xrightarrow{SSB}me_\mu^a+\sigma_\mu^a,\qquad\phi^5\xrightarrow{SSB}v+\rho,
\end{equation}
with $\sigma_\mu^{ab}$ and $\sigma_\mu^a$ being the quantum fluctuations of the fields $A_\mu^{ab}$ and $A_\mu^{a5}$ around their corresponding classical background fields $a_\mu^{ab}$ and $a_\mu^a$, respectively, and $\rho$ the quantum fluctuation of the field $\phi^5$ around its v.e.v. $v=\langle\phi^5\rangle$. The quantities $a_\mu^{ab},$ $a_\mu^a$ and $v$ will henceforth be referred to as the ``classical values'' of the fields $A_\mu^{ab}$, $A_\mu^{a5}$ and $\phi^5$ respectively. As a consequence, the dictionary for emergent gravity can now be updated as follows:
\begin{equation}\label{eq:vev-dictionary}
    \begin{split}
    m^{-1}a_\mu^a\,\text{or}\,\pm v^{-1}m^{-1}\nabla_\mu\phi^A&\xrightarrow{SSB}e_\mu^a,\\
    4vmJ^{-1}w_A^\mu&\xrightarrow{SSB}e_a^\mu,\\
    -\frac{1}{24}v^{-5}m^{-4}J&\xrightarrow{SSB}e=\sqrt{-g},\\
    m^{-2}\eta_{ab}a_\mu^aa_\nu^b\,\text{or}\,v^{-2}m^{-2}P_{\mu\nu}&\xrightarrow{SSB}\eta_{ab}e_\mu^ae_\nu^b\equiv g_{\mu\nu},\\
    16v^2m^2J^{-2}\eta^{AB}w_A^\mu w_B^\nu&\xrightarrow{SSB}\eta^{ab}e_a^\mu e_b^\nu\equiv g^{\mu\nu},
    \end{split}
\end{equation}
with \emph{only} the classical values of the fields $A_\mu^{AB}$ and $\phi^A$ appearing on the left-hand side of these expressions.\\

Identifying a part of the effective geometric quantities as the one arising from considering only the classical values of the fields $A_\mu^{AB}$ and $\phi^A$ implies that such quantities receive contributions from the fluctuations of those fields too, so that each `total' effective geometric quantity can now be decomposed as the sum of a `classical' and a `quantum' contribution, respectively:
\begin{equation}
    \begin{split}
    \mathcal{E}_\mu^a&\equiv e_\mu^a+\hat{e}_\mu^a,\qquad\mathcal{E}_a^\mu\equiv e^\mu_a+\hat{e}^\mu_a,\qquad\mathcal{E}\equiv e+\hat{e},\\
    \mathcal{G}_{\mu\nu}&\equiv g_{\mu\nu}+\hat{g}_{\mu\nu},\qquad\mathcal{G}^{\mu\nu}\equiv g^{\mu\nu}+\hat{g}^{\mu\nu}.
    \end{split}
\end{equation}
Specifically, the classical values of the pre-geometric fields are distinguished from their quantum counterparts in that the former and only the former are subject to geometric conditions after the SSB. In particular, the effective tetrad fields must satisfy the soldered property of spacetime (Eq. \eqref{eq:soldered}) and the tetrad postulate \cite{capozziello:2022zzh}. The quantum parts, on the other hand, are totally unconstrained and, thus, unrelated to the emerging classical background if not as an \emph{effective} description. For example, we have that the quantum part of the effective tetrad is given by $\hat{e}_\mu^a\xleftarrow{SSB}m^{-1}\sigma_\mu^a$, while the quantum part of the effective metric is given by
\begin{equation}
    \hat{g}_{\mu\nu}=\mathcal{G}_{\mu\nu}-g_{\mu\nu}\xleftarrow{SSB}m^{-2}\eta_{AB}A_\mu^{A5}A_\nu^{B5}-m^{-2}\eta_{ab}a_\mu^aa_\nu^b=m^{-2}\eta_{ab}(\sigma_\mu^a\sigma_\nu^b+2a_{(\mu}^a\sigma_{\nu)}^b).
\end{equation}
In light of these considerations, from now on, the classical part of each effective geometric quantity in the unbroken phase will be taken to be the corresponding \emph{fundamental} geometric quantity in the spontaneously broken phase, according to the dictionary \eqref{eq:vev-dictionary}. For instance, the classical part $g_{\mu\nu}$ of the total effective metric $\mathcal{G}_{\mu\nu}$, which emerges after the SSB of the quantity $m^{-2}\eta_{ab}a_\mu^aa_\nu^b$, will be assumed to be the \emph{actual} metric obeying the laws of an emergent metric theory of gravity, that is the Einstein--Cartan theory, and not a mere combination of the pre-geometric fields $A_\mu^{AB}$ or $\phi^A$. In other words, $g_{\mu\nu}$ satisfies the Einstein equations, while $\hat{g}_{\mu\nu}$ does not. This will allow us to use the classical part of the effective metric to raise and lower spacetime indices and to construct adequate Lagrangian densities for a complete theory of emergent gravity coupled to matter.\\

The overall picture of this formalism is that the total effective geometric quantities are
\begin{equation}\label{eq:complete_dictionary}
    \begin{split}
    m^{-1}A_\mu^{a5}\,\text{or}\,\pm v^{-1}m^{-1}\nabla_\mu\phi^A&\xrightarrow{SSB}\mathcal{E}_\mu^a,\\
    4vmJ^{-1}w_A^\mu&\xrightarrow{SSB}\mathcal{E}_a^\mu,\\
    -\frac{1}{24}v^{-5}m^{-4}J&\xrightarrow{SSB}\mathcal{E}=\sqrt{-\mathcal{G}},\\
    m^{-2}\eta_{AB}A_\mu^{A5}A_\nu^{B5}\,\text{or}\,v^{-2}m^{-2}P_{\mu\nu}&\xrightarrow{SSB}\eta_{ab}\mathcal{E}_\mu^a\mathcal{E}_\nu^b\equiv\mathcal{G}_{\mu\nu},\\
    16v^2m^2J^{-2}\eta^{AB}w_A^\mu w_B^\nu&\xrightarrow{SSB}\eta^{ab}\mathcal{E}_a^\mu\mathcal{E}_b^\nu\equiv\mathcal{G}^{\mu\nu},
    \end{split}
\end{equation}
with their quantum parts surviving in a theory of emergent gravity as interactions of the fluctuations of the fields $A_\mu^{AB}$ or $\phi^A$ with themselves or with the matter fields on a classical dynamical background described by $g_{\mu\nu}$.\\

In this formulation, the graviton, meant as the alleged massless spin-$2$ particle that mediates the gravitational interaction in quantum field theory, can be naturally recovered from the quantisation of the classical metric $g_{\mu\nu}$, with the quantum counterpart $\hat{g}_{\mu\nu}$ of the total effective metric being nothing but a rewriting of interactions of the fluctuations of the pre-geometric fields in the spontaneously broken phase. Thus $\hat{g}_{\mu\nu}$ serves only as an \emph{effective} description. This is partly correct, in the sense that the pre-geometric fields $A_\mu^{AB}$ and $\phi^A$ are indeed responsible for determining a classical metric structure in the spontaneously broken phase. Nonetheless, this is ultimately devoid of meaning at the fundamental level of the quantum theory, since the fields $A_\mu^{AB}$ have spin $1$ rather than $2$.

\section{Matter couplings}\label{sec:6}
\subsection{Effective kinetic terms for matter fields}
Armed with a dictionary to translate between pre-geometric quantities in the unbroken phase and effective geometric quantities in the spontaneously broken phase, we are now ready to discuss the implementation of matter fields in a theory of emergent gravity. The criterion that we will abide by in constructing the effective kinetic terms for the matter fields in the unbroken phase is a form of the {\it correspondence principle}: the pre-geometric kinetic terms must reproduce, in the spontaneously broken phase, the correct generally covariant kinetic terms of quantum field theory, when the classical values only of the pre-geometric fields $A_\mu^{AB}$ and $\phi^A$ are considered, so that all classical geometric quantities like $\sqrt{-g}$ and $g^{\mu\nu}$ are well defined. All contributions coming from the quantum part of the effective geometric quantities in the effective kinetic terms will then have to be considered as additional interactions between the fluctuations $\sigma_\mu^{ab}$, $\sigma_\mu^a$ and $\rho$ of $A_\mu^{ab}$, $A_\mu^{a5}$ and $\phi^5$ respectively and the matter fields.\\

The unique form of the effective kinetic terms in the unbroken phase for spacetime scalar ($\Phi$), spinor\footnote{To suitably define the Clifford algebra in the unbroken phase, a fifth gamma matrix must be introduced (see Ref.\ \cite{westman:cartan} for the details).} ($\Psi$) and Yang--Mills ($G$) fields that complies with the correspondence principle can thus be reconstructed using a reverse engineering method. The results are respectively
\begin{subequations}
    \begin{align}
    \mathcal{L}_\Phi=&\frac{1}{3}v^{-3}m^{-2}J^{-1}w_A^\mu w_B^\nu\eta^{AB}\partial_\mu\Phi\partial_\nu\Phi\xrightarrow{SSB}-\frac{1}{2}\sqrt{-g}g^{\mu\nu}\partial_\mu\Phi\partial_\nu\Phi\nonumber\\
    \text{or}\qquad \,\mathcal{L}_\Phi=&-\frac{1}{288}v^{-11}m^{-10}J\epsilon^{\mu\alpha\gamma\rho}\epsilon^{\nu\beta\delta\sigma}P_{\alpha\beta}P_{\gamma\delta}P_{\rho\sigma}\partial_\mu\Phi\partial_\nu\Phi\nonumber\\
    &\xrightarrow{SSB}\frac{1}{12}\sqrt{-g}\epsilon^{\mu\alpha\gamma\rho}\epsilon^{\nu\beta\delta\sigma}g_{\alpha\beta}g_{\gamma\delta}g_{\rho\sigma}\partial_\mu\Phi\partial_\nu\Phi,\label{eq:L_Phi}\\
    \mathcal{L}_\Psi=&-\frac{i}{6}v^{-4}m^{-3}\bar{\Psi}w_A^\mu\gamma^A\nabla_\mu\Psi\xrightarrow{SSB}i\sqrt{-g}\bar{\Psi}\gamma^\mu\nabla_\mu\Psi,\\
    \mathcal{L}_G=&\frac{8}{3}v^{-1}J^{-3}w_A^\rho w_B^\mu w_C^\sigma w_D^\nu\eta^{AB}\eta^{CD}\trace(G_{\mu\nu}G_{\rho\sigma})\xrightarrow{SSB}-\frac{1}{4}\sqrt{-g}g^{\rho\mu}g^{\sigma\nu}\trace(G_{\mu\nu}G_{\rho\sigma})\nonumber\\
    \text{or}\qquad \,\mathcal{L}_G=&-\frac{1}{384}v^{-9}m^{-8}J\epsilon^{\mu\nu\alpha\beta}\epsilon^{\rho\sigma\gamma\delta}P_{\alpha\gamma}P_{\beta\delta}\trace(G_{\mu\nu}G_{\rho\sigma})\nonumber\\
    &\xrightarrow{SSB}\frac{1}{16}\sqrt{-g}\epsilon^{\mu\nu\alpha\beta}\epsilon^{\rho\sigma\gamma\delta}g_{\alpha\gamma}g_{\beta\delta}\trace(G_{\mu\nu}G_{\rho\sigma}),\label{eq:L_G}
    \end{align}
\end{subequations}
where only the classical values of the fields $A_\mu^{AB}$ and $\phi^A$ were considered on the left-hand side of these expressions. Each of the effective kinetic terms for the matter fields in the unbroken phase gives rise, through the quantum part of the effective geometric quantities, to potential terms other than the correct kinetic term in the spontaneously broken phase. The Klein--Gordon and YM Lagrangian densities were both expressed in two different forms, with the second one in both cases containing the contravariant Levi-Civita symbol and the spacetime metric in place of the inverse of the spacetime metric. While these two types of forms are equivalent in the Minkowski spacetime, they differ at a perturbative level and thus also at a quantum level. Nonetheless, we have written them down for completeness.\\

Let us observe that the effective kinetic terms are independent of the mass dimension of the field $\phi^A$, because to each factor of $\phi^A$ or $\nabla_\mu\phi^A$ corresponds a factor $v^{-1}$. Furthermore, all the effective kinetic terms, at least in one of the possible forms, are polynomial in the pre-geometric fields. The term $\mathcal{L}_G$ is the only one which, in one of the two possible forms, does not have a negative power of the mass parameter $m$ as a multiplicative factor and so, apparently, the only one which could be perturbatively renormalisable in the unbroken phase.\\

The formalism under consideration shows no obstruction in reconstructing all the potential terms of the Lagrangian density of the Standard Model of particles, including those for the generation of masses. Indeed, it is sufficient to recast them in the general form $\mathcal{L}_\textup{int}=-\lvert J\rvert V$, where $V$ is a functional of the matter fields in the Minkowski spacetime, and substitute the Minkowski metric with the expression for the effective metric given in the dictionary \eqref{eq:complete_dictionary} for those interactions that require it. In addition to that, the gauge principle for matter fields is unrelated to the background geometry of spacetime, hence it can be effortlessly implemented too. Therefore, the Standard Model of particles can be safely embedded in the formalism of a theory of emergent gravity, even in the unbroken phase where spacetime is devoid of a metric structure.

\subsection{Classical vs.\ quantum equivalence principle}
The most notable prediction that follows from the form of the effective kinetic terms in the unbroken phase is arguably the following: to correctly reproduce quantum field theory on curved spacetime (based on the implementation of GR's EEP into the formalism of the quantum theory), the surviving quantum fluctuations of the pre-geometric fields must interact with the various matter fields in a non-universal way in the spontaneously broken phase. This can be interpreted as a violation of the EEP due to the quantum part of the spacetime geometry. In other words, it is exactly imposing the EEP at a classical, geometric level that in this construction leads to the violation of such principle at the quantum level. Since each effective kinetic term is suppressed by a negative power of $v$, the violation of the EEP at the quantum level can remain sufficiently small in the spontaneously broken phase, not to be detectable in experimentally tested regimes up to now, so that the EEP can still be said to hold true in the classical limit. This is actually an important reason to argue in favour of the assumption discussed in Sec.\ \ref{sec:3} about $k_\textup{MM}$ and $k_\textup{W}$ being relatively small and $v$ being relatively big, because otherwise quantum violations of the EEP would be observable at lower energies. This suggests that these theories predict violations of the EEP at a microscopic level, while any deviations at astrophysical scales are highly suppressed as the quantum fluctuations of the pre-geometric fields become negligible.\\

In any case, to be more thorough, it must be remarked that an actual implementation of the EEP requires not only a recovery of the correct form of the kinetic terms for matter, but also their relative numerical coefficients to be correct. In fact, the local Lorentz invariance of spacetime is a necessary but not sufficient condition for the EEP to hold. The other necessary condition is the validity of the weak equivalence principle (WEP), also known as the universality of free fall, which assumes the universality of the coupling between the spacetime geometry and every matter field. Thus, while the proposed theory of emergent gravity recovers the EEP in the gravitational sector, it remains `agnostic' about it when the matter sector is introduced, because it cannot predict the relative numerical coefficients for the matter couplings. In other words, strictly speaking this theory does not predict the validity of the WEP, but neither forbids it. So the EEP can very simply be accommodated in the formalism at a classical level, but that is not a necessity. This motivates experimental precision tests of the WEP and the E\"{o}tv\"{o}s ratio $\eta$ (see Refs.\ \cite{will:confrontation,tino:precision} and references therein). Such parameter is defined as
\begin{equation*}
    \eta\equiv2\frac{\lvert a_1-a_2\rvert}{\lvert a_1+a_2\rvert}=2\frac{\Bigl\lvert m_1^{(g)}/m_1^{(i)}-m_2^{(g)}/m_2^{(i)}\Bigr\rvert}{\Bigl\lvert m_1^{(g)}/m_1^{(i)}+m_2^{(g)}/m_2^{(i)}\Bigr\rvert}
\end{equation*}
and measures the fractional difference in the acceleration $a$ between two freely falling bodies or, equivalently, the fractional difference in the ratios of their gravitational mass $m^{(g)}$ and inertial mass $m^{(i)}$. For other theories that predict a violation of the equivalence principle in various ways, consult the Refs.\ \cite{damour:theoretical,damour:dilaton,orlando:superpositions,adunas:probing}. The upshot of this analysis is then that the EEP can be emergent from a fundamental gauge principle formulated on a pre-geometric spacetime via a mechanism of SSB and that, if so, its quantum violations could be highly suppressed. This conclusion hinges on the validity of the WEP, which is not guaranteed and can be experimentally tested to high precision.

\section{Kinematics of pre-geometric fields}\label{sec:7}
In this section we discuss the construction of the effective kinetic terms for the fields $A_\mu^{AB}$ and $\phi^A$ in the unbroken phase. We start by observing that, without assuming the existence of a spacetime metric, the very notion of propagation, as commonly understood in terms of kinetic terms and propagators, could be counter-intuitive, both for pre-geometric fields and matter fields. In purely mathematical terms, however, the notions of derivation and integration on a differentiable manifold are more primitive than that of a metric, and so they can still provide a consistent formulation of the action principle and of the propagation of fields within this framework. Moreover, $n$-point field correlators can be defined and derived from the path integral approach -- even if Wick's theorem will not grant them a decomposition in terms of standard propagators. We reserve a more in-depth mathematical study of these aspects for future works.\\

The form of $\mathcal{L}_A$ is the same as that of the effective kinetic term for YM fields shown in the Eq.\ \eqref{eq:L_G}. The form of $\mathcal{L}_\phi$, instead, demands more attention, mainly because of the fact that the mass dimension of the field $\phi^A$ has so far remained unspecified. Since the mass dimensions of all coupling constants are \emph{inversely} proportional to some power of the mass dimension of the field $\phi^A$, inspection of the complete Lagrangian density $\mathcal{L}$ reveals that the theory is non-renormalisable for any positive value of the mass dimension of $\phi^A$. Thus, the only way
 to have dimensionless coupling constants is for the field $\phi^A$ itself to be dimensionless, that is $[\phi]=[M]^0$. We will set out the consequences of the latter possibility, inclusive of both advantages and complications.\footnote{The idea that $[\phi]=[M]^0$ is originally due to {\it Wilczek}, who presented it without any explicit arguments \cite{wilczek:gauge}. We assume that {\it Wilczek}'s choice is motivated by the potential power-counting renormalisability of dimensionless coupling constants.} The main advantage is, of course, highly intriguing: the (free) theory is potentially power-counting renormalisable in the unbroken phase. This rationale is in agreement with the fact that what matters in Quantum Field Theory are not really the `absolute' mass dimensions of fields, but their `\emph{relative}' mass dimensions, i.e.\ relative to those obtained by choosing a dimensionless constant in front of the kinetic terms ($-1/2$ for scalars and tensors, $i$ for spinors, $-1/4$ for vectors). This is a choice motivated by the requirement that the free theory (that is, \emph{at least} the free theory) is renormalisable.\\

Following the expression \eqref{eq:L_Phi}, the appropriate effective kinetic term for a dimensionless spacetime scalar, internal-space vector $\phi^A$ is then
\begin{equation}\label{eq:kinetic-phi}
    \mathcal{L}_\phi=\frac{1}{3}v^{-3}J^{-1}w_A^\mu w_B^\nu\eta^{AB}\eta_{CD}\nabla_\mu\phi^C\nabla_\nu\phi^D\xrightarrow{SSB}-\frac{1}{2}m^2\sqrt{-g}g^{\mu\nu}\eta_{AB}\nabla_\mu\phi^A\nabla_\nu\phi^B,
\end{equation}
where the classical values only of the pre-geometric fields were considered on the left-hand side of this expression. This result is a reminder that the Klein--Gordon Lagrangian density for a free \emph{dimensionless} scalar field $\Phi$ with mass $m_\Phi$ in quantum field theory on curved spacetime must be recast as
\begin{equation}
    \mathcal{L}'_\textup{KG}=-\frac{1}{2}m^2\sqrt{-g}g^{\mu\nu}\partial_\mu\Phi\partial_\nu\Phi-\frac{1}{2}m^2\sqrt{-g}m_\Phi^2\Phi^2,
\end{equation}
with an overall factor $m^2$ that does not affect the equation of motion.\\

As a side note, we observe that it is possible to contract the internal-space indices of the effective kinetic term for $\phi^A$ in the unbroken phase in another way, namely
\begin{equation*}
    \frac{2}{3}v^{-3}J^{-1}w_A^\mu w_B^\nu\nabla_\mu\phi^{(A}\nabla_\nu\phi^{B)}.
\end{equation*}
Nonetheless this term will be neglected on the basis that, in the spontaneously broken phase, it does not lead to a term which is quadratic in the partial derivatives of $\rho$. Instead, it simply reduces to an additional interaction term for the pre-geometric fields.

\section{The Higgs mechanism for gravity}\label{sec:8}
Our complete Lagrangian density is initially the sum of the following terms: the gravitational terms $\mathcal{L}_\textup{MM}$ or $\mathcal{L}_\textup{W}$ to recover classical gravity that are both retained at first; the effective kinetic terms $\mathcal{L}_A$ and $\mathcal{L}_\phi$ for the kinematics of the pre-geometric fields; the potential represented by (the opposite of) $\mathcal{L}_\textup{SSB}$ to realise the mechanism of SSB; the effective kinetic terms $\mathcal{L}_\Phi$, $\mathcal{L}_\Psi$ and $\mathcal{L}_G$ for the coupling of matter fields to gravity; the generalisation $\mathcal{L}_\textup{int}$ of all the interactions of the Standard Model of particles. Therefore, the total Lagrangian sums up to 
\begin{equation}\label{eq:lagrangian}
    \mathcal{L}=\underbrace{\mathcal{L}_\textup{MM}+\mathcal{L}_\textup{W}}_{\substack{\text{gravitational}\\\text{sector}}}+\underbrace{\mathcal{L}_A+\mathcal{L}_\phi}_{\substack{\text{kinematics of}\\\text{pre-geometric}\\\text{fields}}}+\underbrace{\mathcal{L}_\textup{SSB}}_{\substack{\text{symmetry-}\\\text{breaking}\\\text{potential}}}+\underbrace{\mathcal{L}_\Phi+\mathcal{L}_\Psi+\mathcal{L}_G}_{\substack{\text{kinematics of}\\\text{matter fields}}}+\underbrace{\mathcal{L}_\textup{int}}_{\substack{\text{interactions of}\\\text{matter fields}}}.
\end{equation}
Note that the complete Lagrangian density is free of Ostrogradsky instabilities in the unbroken phase.\\

The Higgs mechanism realised by the field $\phi^A$ after the SSB from the $SO(1,4)$ or $SO(3,2)$ gauge group (with ten generators) to the $SO(1,3)$ gauge group (with six generators) allows for the ``eating up'' mechanism of the four Goldstone bosons $\phi^A$ with $A\ne5$ and gives mass, for each value of the covariant spacetime index, to the four fluctuations $\sigma_\mu^a$ of the gauge fields $A_\mu^{a5}$, while keeping the fluctuations $\sigma_\mu^{ab}$ of the gauge fields $A_\mu^{ab}$ massless.\\

The surviving pre-geometric fields in the spontaneously broken phase can in principle suffer from two kinds of instabilities: Ostrogradsky instabilities give rise to ghosts because of a kinetic term with the wrong sign \cite{woodard:ostrogradsky}, while tachyonic instabilities are due to a mass term with the wrong sign. Specifically, in this section we will also show under which circumstances the complete Lagrangian density \eqref{eq:lagrangian} can have two possible instabilities, namely the field $\rho$ could be a ghost and some of the fields $\sigma_\mu^a$ could be tachyons. In particular, we expect some of the fields $\sigma_\mu^a$ to be tachyons because of the Lorentzian signature of the spacetime and internal-space metrics. Indeed, this can be seen from inspecting the form of the Proca Lagrangian density for some vector fields $B_\mu^{a}$ with mass $m_B$ and gauge group $SO(1,4)/SO(1,3)$:
\begin{equation}
    \mathcal{L}_\textup{P}=-\frac{1}{4}\sqrt{-g}g^{\rho\mu}g^{\sigma\nu}\trace(G_{\mu\nu}G_{\rho\sigma})-\frac{1}{2}\sqrt{-g}g^{\mu\nu}m_B^2\eta_{ab}B_\mu^aB_\nu^b,
\end{equation}
where the mass term can be expanded more explicitly as
\begin{equation*}
    -\frac{1}{2}\sqrt{-g}g^{\mu\nu}m_B^2(-B_\mu^1B_\nu^1+B_\mu^2B_\nu^2+B_\mu^3B_\nu^3+B_\mu^4B_\nu^4).
\end{equation*}
This issue is ultimately due to the non-compactness of the gauge group and can be alternatively stated in terms of states with negative norm. For more details on a consistent quantisation with non-compact gauge groups, see the Refs.\ \cite{margolin:sigma,margolin:yang-mills}. In any case, tachyonic instabilities hinge crucially on the form of the interactions and the self-interactions of tachyonic fields, hence we will proceed to study them in an unbiased way in the following analysis.\\

To examine the main results of the Higgs mechanism, we will expand the complete Lagrangian density \eqref{eq:lagrangian} up to the second order in the fluctuations of the pre-geometric fields. The details of the computations are presented in the App.\ \ref{sec:appendix_B}.

\subsection{$\Lambda$ sector}
The contribution to the cosmological constant coming from the $0$\textsuperscript{th} order expansion of $\mathcal{L}_\phi$, i.e.\ $-2v^2m^4\sqrt{-g}$, has a drastically different value depending on whether it is computed from $\mathcal{L}_\textup{MM}$ or $\mathcal{L}_\textup{W}$. In the former case, one obtains a dominant contribution of order $\mathcal{O}(v)$ (unless $\lvert k_\textup{MM}\rvert\gg10^{60}$). According to cosmological observations, this would be unacceptable, as the Lagrangian density for the effective cosmological constant becomes
\begin{equation}
    \mathcal{L}_\Lambda\equiv-\biggl(\pm3m^2\pm\frac{vm^2}{16k_\textup{MM}}\biggr)M_\textup{P}^2\sqrt{-g}.
\end{equation}
In the latter case, instead, one obtains a sub-leading contribution of the order $\mathcal{O}(v^{-1})$ (unless $\lvert k_\textup{W}\rvert\ll10^{-58}$). This can very well be within experimental bounds, as
\begin{equation}
    \mathcal{L}_\Lambda\equiv-\biggl(\pm6m^2-\frac{m^2}{4k_\textup{W}v}\biggr)M_\textup{P}^2\sqrt{-g}.
\end{equation}
This result provides a strong argument for rejecting $\mathcal{L}_\textup{MM}$ and realising the gravitational sector with $\mathcal{L}_\textup{W}$ only in what follows.

\subsection{$\rho$ sector}
Collecting all terms containing only the fluctuation $\rho$ up to the second order, one finds
\begin{equation}
    \begin{split}
    \mathcal{L}_\rho&\equiv\mp\frac{1}{2}m^2\sqrt{-g}g^{\mu\nu}\partial_\mu\rho\partial_\nu\rho+\sqrt{-g}[12k_\textup{W}v^2m^2(-e_a^\mu e_b^\nu R_{\mu\nu}^{ab}\pm12m^2)-4vm^4]\rho\\
    &+\sqrt{-g}[12k_\textup{W}vm^2(-e_a^\mu e_b^\nu R_{\mu\nu}^{ab}\pm12m^2)-96k_\textup{SSB}v^3m^4-2m^4]\rho^2.
    \end{split}
\end{equation}
Since the kinetic term has the wrong sign in the case of $SO(3,2)$, the field $\rho$ is a ghost if the fundamental gauge group of the theory is taken to be the anti-de Sitter group. This is one reason to prefer $SO(1,4)$ to $SO(3,2)$ in our construction. Thus, in the case of $SO(1,4)$, the bare mass of the field $\rho$ is given by
\begin{equation}\label{eq:mass_rho}
    \begin{split}
    m_\rho^2&=192k_\textup{SSB}v^3m^2-288k_\textup{W}vm^2+4m^2=-24\frac{k_\textup{SSB}}{k_\textup{W}}M_\textup{P}^2+\frac{36}{v^2}M_\textup{P}^2+\frac{2}{3}\Lambda\\
    &\approx(1.43\cdot10^{38}k_\textup{SSB}+2.07\cdot10^{-42}+2.83\cdot10^{-84})\,\text{GeV}^2.
    \end{split}
\end{equation}
Observe that the contribution to $m_\rho$ coming from the gravitational sector is much bigger than the one coming from the effective kinetic term, while the contribution of the symmetry-breaking potential depends on the free parameter $k_\textup{SSB}$ and is even higher unless $k_\textup{SSB}\ll10^{-80}$; in particular, one finds that $m_\rho\sim M_\textup{P}$ for $k_\textup{SSB}\sim10^{-2}$. Another possibility for having $m_\rho\sim M_\textup{P}$, but with the leading contribution coming from the gravitational sector rather than the symmetry-breaking potential, is realised if $v\sim10$ whenever $k_\textup{SSB}\ll10^{116}$.

\subsection{$\sigma_\mu^a$ sector}
In a similar fashion, collecting all terms containing only the fluctuations $\sigma_\mu^a$ up to the second order, one obtains
\begin{equation}
    \begin{split}
    \mathcal{L}_\sigma&\equiv-\frac{1}{4}\sqrt{-g}g^{\rho\mu}g^{\sigma\nu}\trace(F_{\mu\nu}F_{\rho\sigma})+2k_\textup{W}v^3m\epsilon_{abcd}\epsilon^{\mu\nu\rho\sigma}(R_{\mu\nu}^{ab}\mp4m^2e_\mu^ae_\nu^b)e_\rho^c\sigma_\sigma^d\\
    &+\biggl[k_\textup{W}v^3\epsilon_{abcd}\epsilon^{\mu\nu\rho\sigma}(R_{\mu\nu}^{ab}\mp12m^2e_\mu^ae_\nu^b)-\frac{1}{2}\sqrt{-g}g^{\rho\sigma}v^2m^2\eta_{cd}\biggr]\sigma_\rho^c\sigma_\sigma^d.
    \end{split}
\end{equation}
Note that every self-interaction term of this Lagrangian density, except the mass term, cannot be expressed in the usual metric, background-independent form -- this is due, in particular, to the absence of the factor $\sqrt{-g}$. In any case, coming from the gravitational sector, such terms are relevant only at the Planck scale, signalling somehow at such scale a foreseeable departure from metricity in the quantum theory. As expected, $\mathcal{L}_\sigma$ describes six tachyons, which are specifically the fields with the spacetime index corresponding to the time direction and the internal-space index corresponding to a spatial direction or vice versa. The bare mass of the physical fields $\sigma_\mu^a$ is given by
\begin{equation}
    m_\sigma^2=v^2m^2=\frac{v^2}{6}\Lambda\approx7.31\cdot10^{-5}\,\text{GeV}^2,
\end{equation}
which is a mesoscale energy as $\Lambda\ll m_\sigma^2\ll M_\textup{P}^2$ (as long as $10^{-60}\ll\lvert k_\textup{W}\rvert\ll10^{120}$). As can be read from the expression of $\mathcal{L}_\sigma$, the self-interaction terms of the second order in $\sigma_\mu^a$ have a coupling constant which is of the order of the Planck scale, meaning that the effective mass of both `ordinary' and tachyonic fields $\sigma_\mu^a$ is Planckian. More precisely, by diagonalising the effective mass matrix for the fields $\sigma_\mu^a$ in the Minkowski spacetime, one obtains four different eigenvalues which are $(c_1M_\textup{P}^2\pm c_2m_\sigma^2)$ and $-(c_1M_\textup{P}^2\pm c_2m_\sigma^2)$,
where $c_1$ and $c_2$ are numerical coefficients that can eventually include quantum corrections.\\

At this point, it is fitting to comment on a peculiar result of the theory: as expected from the Higgs mechanism, the quantum part $\hat{e}_\mu^a$ of the total effective tetrad $\mathcal{E}_\mu^a$ becomes massive, but surprisingly its classical part $e_\mu^a$ does not. This is because the possible mass term for $e_\mu^a$ is exactly what yields the quantum correction to the tree-level value of the cosmological constant, i.e.\ the term $-2v^2m^4\sqrt{-g}$ discussed above. Although unexpected, this is a welcome result, for it guarantees that gravitational waves (computed from the \emph{classical} metric $g_{\mu\nu}=\eta_{ab}e_\mu^ae_\nu^b$ as wavelike perturbations with respect to the Minkowski spacetime) have the same number of degrees of freedom as predicted by GR.

\subsection{$\sigma_\mu^{ab}$ sector}
Finally, the fields $\sigma_\mu^{ab}$ remain massless, as can be verified in the App.\ \ref{sec:appendix_B}. This is an important consistency check, since after the SSB it is not the tetrads that behave as gauge potentials in the YM interpretation of the theory, but rather the spin connection. Therefore, the Higgs mechanism for gravity presented in this section accomplishes, via a symmetry-breaking field $\phi^A$, the goal of making contact between the fundamental gravitational and quantum theories by recovering both the tetrads of gravity and massless gauge potentials {\it à la} Yang--Mills out of a \emph{single} pre-geometric gauge field $A_\mu^{AB}$.

\section{Cosmological implications: from a pre-geometric to a metric Universe}\label{sec:9}
Our best cosmological model for the evolution of the Universe is based on GR and the Standard Model of particles. The consequences of restoring the fundamental gauge symmetry of spacetime in the very early Universe are therefore radical in the context of a theory of emergent gravity. In particular, the pre-geometric character of spacetime in the unbroken phase could point at a few different ways to solve the problem of the Big Bang singularity. The most straightforward interpretation of the theory is that of an eternal pre-geometric universe (one in which the notion of spacetime distance is not even defined), which undergoes a phase transition that spontaneously breaks the fundamental gauge symmetry of spacetime and leads to the emergence of an expanding cosmos as commonly understood in the geometric sense. The notion of metric distance would simply cease to make sense at the Planck scale, where the unbroken phase of the theory is possibly restored. Another intriguing possibility is that of recovering a version of conformal cyclic cosmology \cite{penrose:outrageous}, if conformal invariance is assumed to be an additional fundamental gauge symmetry of spacetime. This could be achieved by extending the gauge group of the present analysis from $SO(1,4)$ or $SO(3,2)$ to $SO(2,4)$, as will be discussed in the next section.\footnote{In this scenario, one would have conformal invariance to begin with, as a fundamental property of the Universe in the unbroken phase, and would not need it to be recovered in a physical way assuming -- quite unsatisfactorily -- that photons are the only particles left at the end of each cosmic eon (see Ref.\ \cite{penrose:outrageous}).} Still another possibility is that the dynamics of bouncing cosmologies is realised from four-fermion repulsive interactions in Einstein--Cartan gravity, in turn obtained after the SSB from $SO(1,4)$ to $SO(1,3)$ \cite{alexander:2014eva,alexander:2014uaa,addazi:2016rnz,addazi:2018zjv} -- while for four-fermion inflation, see the Ref.\ \cite{addazi:2017qus}.\\

Based on rather general arguments of thermal field theory -- see, for example, the Refs.\ \cite{mukhanov:cosmology,quiros:temperature} -- the mechanism of SSB from either $SO(1,4)$ or $SO(3,2)$ to $SO(1,3)$ happens at a critical temperature $T_c\sim m_\rho$, inducing a phase transition from a pre-geometric spacetime to one that expands according to the metric laws governing the evolution of its emergent curvature. For this reason, one would expect the free parameter characterising the symmetry-breaking potential $-\mathcal{L}_\textup{SSB}$ to be $k_\textup{SSB}\lesssim10^{-2}$ (if $v\sim10^{40}$) in order to have $T_c\lesssim M_\textup{P}$. Note that even if the field $\phi^A$ is dimensionless, there is no issue in understanding the definition of all physical observables. In particular, its v.e.v. $v$ enters the predictions for the physical quantities by simply providing a normalisation factor in terms of the ratio between $M_\textup{P}^2$ and $\Lambda$.\\

An attractive possibility is that the SSB is related to a primordial first-order phase transition (FOPT). In this case, we speculate that it could source gravitational waves in the form of a stochastic background with ultra-high frequencies, i.e.\ the echo of what could be called ``metro-genesis'' (assuming that it can survive a `washing out' due to a possible inflationary scenario). In principle, the study of this phenomenon involves thermal field theory analyses that are not yet fully developed and understood in the context of this theory. However, typically the gravitational wave frequency peak is directly proportional to the critical temperature of enucleation. For example, an electroweak FOPT corresponds to space-based interferometer frequency windows, around the mHz. A GHz window can instead arise to test ultra-high energy FOPTs up to the Planck domain \cite{addazi:2018nzm}. This range represents the next frontier for several experiments such as resonant cavities, table-top experiments and graviton-photon radio astronomy \cite{addazi:2024osi,addazi:2024kbq}.\\

If the cosmic history, in the metric sense, begins at $T_c$, it is then natural to wonder about the role of time in the prehistoric Universe, that is the pre-geometric spacetime of the unbroken phase. Taking the mathematics at face value, one is led to think that time and evolution in time are well-defined concepts in the prehistoric Universe too, as the former is simply one of the four coordinate labels of the spacetime manifold, and the latter is characterised by the partial derivative with respect to time of any given tensor field. What actually goes missing in the absence of a spacetime metric is the notion of causality, as the norm of four-vectors ceases to be defined. Even if the coordinates of the spacetime manifold are well-defined, the corresponding four-position vectors do not have a well-defined norm (at least not in the ordinary way), hence the notions of spacetime interval, light cone structures, causality and microcausality simply fail in the prehistoric Universe. The SSB of the theory of emergent gravity under consideration can therefore be said to be \emph{acausal} and, as a consequence, the physical quantum of the Higgs-like field $\phi^A$ can be regarded as a {\it bona fide} ``God particle''.

\section{Minimal extension for conformal symmetry}\label{sec:10}
Out of the four hypotheses discussed in Subsec.\ \ref{subsec:3.5} for the uniqueness of the Lagrangian density of a consistent theory of emergent gravity, the fourth is the only one that can be safely altered without running into trouble.\footnote{In Ref.\ \cite{Zlosnik:2016fit}, the authors attempted the construction of a conformal gauge theory of gravity based on the $SU(2,2)$ gauge group.} A minimal extension of the pre-geometric field content is given by the introduction of a second spacetime scalar, internal-space vector field $\Tilde{\phi}^{\Tilde{A}}$. These additional degrees of freedom allow for the study of a pre-geometric field theory whose gauge group is a minimal extension of $SO(1,4)$, that is $SO(2,4)$. Since $SO(2,4)$ is locally isomorphic to the conformal group of spacetime, it can be implemented to formulate a theory of emergent gravity with a pre-geometric spacetime whose gauge symmetries include the conformal symmetry in the unbroken phase. The symmetry-breaking pattern of such theory is due to two distinct mechanisms of SSB, first from $SO(2,4)$ to $SO(1,4)$, and then from $SO(1,4)$ to $SO(1,3)$. The second step of SSB is realised by the Higgs-like field $\phi^A$ as described in the previous sections, while the first step, that we will denote as $\widetilde{\text{SSB}}$, is realised by the other Higgs-like field $\Tilde{\phi}^{\tilde{A}}$, as will be discussed in the remaining of this section.\\

The easiest way to extend the Lagrangian density \eqref{eq:lagrangian} to include the symmetries of the gauge group $SO(2,4)$ and the dynamics of an additional (dimensionless) pre-geometric field $\Tilde{\phi}^{\Tilde{A}}$ can be described in three steps. First of all, the new characteristics of the tangent spaces can be implemented by applying the following prescription:
\begin{equation}\label{eq:prescription}
\epsilon_{ABCDE}\rightarrow\frac{1}{\Tilde{v}}\epsilon_{\Tilde{A}\Tilde{B}\Tilde{C}D\Tilde{E}\Tilde{F}}\Tilde{\phi}^{\Tilde{F}},
\end{equation}
with $\Tilde{v}$ a nonzero numerical constant, whose physical meaning will shortly be clarified. Furthermore, in the above expression, the uppercase Latin indices without a tilde run from $1$ to $5$ on the left-hand side, while the uppercase Latin indices with a tilde run from $1$ to $6$ on the right-hand side. Thus, the indices of the internal-space generalised Minkowski metric, $\eta_{AB}$ or $\eta_{\Tilde{A}\Tilde{B}}$, will run from $1$ to $5$ or from $1$ to $6$ depending on whether the gauge group is respectively $SO(1,4)$ or $SO(2,4)$, with signature $(-,-,+,+,+,+)$ in the latter case. Thus, the field multiplets under consideration, respectively $\phi^A$ and $\Tilde{\phi}^{\Tilde{A}}$, have five components in the fundamental representation of $SO(1,4)$ and six components in the fundamental representation of $SO(2,4)$. On the other hand, for the $SO(2,4)$ case, there are fifteen gauge fields $A_\mu^{\Tilde{A}\Tilde{B}}$ for each value of the covariant spacetime index. Furthermore, the field $\Tilde{\phi}^{\Tilde{A}}$ can be made dynamical by adding the following effective kinetic term, which is akin to that of Eq.\ \eqref{eq:kinetic-phi} for $\phi^A$, namely 
\begin{equation}
    \mathcal{L}_{\Tilde{\phi}}=\frac{1}{3}v^{-3}\Tilde{J}^{-1}w_{\Tilde{A}}^\mu w_{\Tilde{B}}^\nu\eta^{\Tilde{A}\Tilde{B}}\eta_{\Tilde{C}\Tilde{D}}\nabla_\mu\Tilde{\phi}^{\Tilde{C}}\nabla_\nu\Tilde{\phi}^{\Tilde{D}},
\end{equation}
where the auxiliary fields $w_{\Tilde{A}}^\mu$ and $\Tilde{J}$ are defined by implementing the prescription \eqref{eq:prescription} into the expressions \eqref{eq:aux_w} and \eqref{eq:aux_J} respectively. The choice of constructing all effective geometric quantities with a single pre-geometric field, $\phi^A$, rather than writing down also quantities with $\Tilde{\phi}^{\Tilde{A}}$ only or `mixed' quantities with both $\phi^A$ and $\Tilde{\phi}^{\Tilde{A}}$, can be understood as aiming to avoid the realisation of multiple metric tensors in the spontaneously broken phase, as it would be the case for bimetric theories of massive gravity, for instance \cite{derham:2010kj}. Finally, the first step of the symmetry-breaking pattern can be realised with a potential for the field $\Tilde{\phi}^{\Tilde{A}}$, which is the analogue of the one expressed in Eq.\ \eqref{eq:potential} for $\phi^A$, namely 
\begin{equation}
    \mathcal{L}_{\widetilde{\textup{SSB}}}=-k_{\widetilde{\textup{SSB}}}\Tilde{v}^{-4}\lvert \Tilde{J}\rvert(\eta_{\Tilde{A}\Tilde{B}}\Tilde{\phi}^{\Tilde{A}}\Tilde{\phi}^{\Tilde{B}}+\Tilde{v}^2)^2,
\end{equation}
with $k_{\widetilde{\textup{SSB}}}$ a positive numerical constant. The potential $-\mathcal{L}_{\widetilde{\textup{SSB}}}$ is minimised for instance when $\Tilde{\phi}^{\Tilde{A}}=\Tilde{v}\delta_2^{\Tilde{A}}$, and any other possible solution for the v.e.v. $\Tilde{v}$ of the field $\Tilde{\phi}^{\Tilde{A}}$ can be put in this form with a gauge transformation. To make sure that this is actually the first and not the second step of the overall symmetry-breaking pattern, one has to assume a specific condition on $\Tilde{v}$ and $v$, which will be referred to as a hierarchy hypothesis in what follows. Therefore, the symmetry-breaking pattern of this theory unfolds as follows: a first Higgs-like field $\Tilde{\phi}^{\Tilde{A}}$ realises the $\widetilde{\text{SSB}}$ from $SO(2,4)$ to $SO(1,4)$, reducing the number of the unbroken generators of the gauge group of spacetime from fifteen to ten. Then, a second Higgs-like field $\phi^A$ realises the SSB from $SO(1,4)$ to $SO(1,3)$, reducing the number of unbroken generators from ten to six. The first phase transition happens at a critical temperature $\Tilde{T}_c$ and singles out, for each value of the covariant spacetime index, the five gauge fields $A_\mu^{\Tilde{A}2}$. These cannot be understood in terms of classical geometric quantities, after the second phase transition that happens at $T_c\ll\Tilde{T}_c$. Moreover, in passing from the unbroken phase with the $SO(2,4)$ gauge symmetry to the intermediate phase with the $SO(1,4)$ gauge symmetry, the internal-space indices need be relabelled so as to run from $1$ to $5$, as this allows to recover exactly the formalism for the second step of SSB as described in the previous sections.\\

The details of the Higgs mechanism for $\widetilde{\text{SSB}}$ followed by SSB are reported in the App.\ \ref{sec:appendix_C}, and lead to a few interesting results. First of all, in the gravitational sector $\mathcal{L}_\textup{W}$ yields a correction $(1+\Tilde{\rho}/\Tilde{v})$ with respect to the expression \eqref{eq:L_W-SSB}, where $\Tilde{\rho}$ is the fluctuation of $\Tilde{\phi}^2$ around its v.e.v. $\Tilde{v}$. Since this correction is linear in $\Tilde{\rho}$, it does not contribute to the mass of such field (unlike what happens for the field $\rho$, see Eq.\ \eqref{eq:mass_rho}). Second, $\mathcal{L}_{\widetilde{\textup{SSB}}}$ yields the same result for $\Tilde{\rho}$ as $\mathcal{L}_\textup{SSB}$ does for $\rho$ (Eq.\ \eqref{eq:L_SSB-SSB}), but with $k_{\widetilde{\textup{SSB}}}$ in place of $k_\textup{SSB}$ and an additional factor $(v/\Tilde{v})^2$. From this, it follows that the correct hierarchy hypothesis is $\Tilde{v}\ll v$ (if $k_{\widetilde{\textup{SSB}}}\sim k_\textup{SSB}$), as it implies $m_{\Tilde{\rho}}\gg m_\rho$ and thus $\Tilde{T}_c\gg T_c$ in terms of critical temperatures. Finally, $\mathcal{L}_{\Tilde{\phi}}$ yields a ghost field $\Tilde{\rho}$ with a mass term for the gauge fields $A_\mu^{\Tilde{A}2}$ that includes tachyons. Therefore, the symmetry-breaking pattern 
$$SO(2,4)\xrightarrow{\widetilde{SSB}}SO(1,4)\xrightarrow{SSB}SO(1,3)$$ 
yields a ghost $\Tilde{\rho}$ and an ordinary scalar field $\rho$.\\

In this setting, the existence of a ghost is inevitable, as can be seen from the fact that the other possible symmetry-breaking pattern, 
$$SO(2,4)\xrightarrow{\widetilde{SSB}}SO(3,2)\xrightarrow{SSB}SO(1,3),$$ yields an ordinary $\Tilde{\rho}$ and a ghost scalar field $\rho$. The physical reason for the existence of a ghost in this framework is that, in breaking the gauge symmetry of the initial group $SO(2,4)$ to that of the final group $SO(1,3)$, two different types of internal-space directions must be broken: one spatial direction and one time direction. Once a signature for the internal-space metric is chosen, in fact, the difference between a spatial and a time direction in each step of the overall SSB is what yields, via the effective kinetic terms, either the correct or the wrong sign for the kinetic terms of the fluctuations of the pre-geometric scalar fields in the spontaneously broken phase.\\

In this section we have analysed the simplest of all theories of emergent gravity that employ the gauge group $SO(2,4)$ and two spacetime scalars, the internal-space vector fields $\phi^A$ and $\Tilde{\phi}^{\Tilde{A}}$. In more complicated theories, involving for example `mixed' effective geometric quantities containing both $\phi^A$ and $\Tilde{\phi}^{\Tilde{A}}$, the results could be qualitatively different. In particular, the hierarchy hypothesis $\Tilde{v}\ll v$ discussed above is clearly model-dependent and it is conceivable that the presence of a ghost scalar field could be avoided (or stabilised), with more elaborate constructions.

\section{Discussion}\label{sec:11}
By requiring formal self-consistency and consistency with cosmological observations, the analysis of the Higgs mechanism conducted in Sec.\ \ref{sec:8} leads to the realisation of a consistent theory of emergent gravity {\it à la} Yang--Mills with a complete Lagrangian density of the form
\begin{equation}
    \mathcal{L}=\underbrace{\mathcal{L}_\textup{W}}_{\substack{\text{gravitational}\\\text{sector}}}+\underbrace{\mathcal{L}_A+\mathcal{L}_\phi}_{\substack{\text{kinematics of}\\\text{pre-geometric}\\\text{fields}}}+\underbrace{\mathcal{L}_\textup{SSB}}_{\substack{\text{symmetry-}\\\text{breaking}\\\text{potential}}}+\underbrace{\mathcal{L}_\Phi+\mathcal{L}_\Psi+\mathcal{L}_G}_{\substack{\text{kinematics of}\\\text{matter fields}}}+\underbrace{\mathcal{L}_\textup{int}}_{\substack{\text{interactions of}\\\text{matter fields}}}
\end{equation}
based on the gauge symmetry of the de Sitter group $SO(1,4)$. The theory $\mathcal{L}_\textup{MM}$ was discarded from the gravitational sector -- in favour of $\mathcal{L}_\textup{W}$ -- because its prediction for the effective value of the cosmological constant is inconsistent with cosmological observations. In addition, the de Sitter group $SO(1,4)$ was preferred to the anti-de Sitter group $SO(3,2)$, on the basis that it provides the right sign for the cosmological constant according to the presently available data and that it does not contain ghosts. From a purely theoretical point of view, it is surprising to prove that a theory grounded on two possible gauge groups, which are equally valid in principle, actually ends up being free of inconsistencies only in the case which is also in agreement with observations.\\

As a matter of fact, this theory of emergent gravity is not a direct recasting of GR as a gauge theory of some sort, but rather a way to recover GR from a gauge theory {\it à la} Yang--Mills via the SSB of its vacuum state. Of course, this reformulation of the gravitational interaction does not remove in any obvious way the issues stemming from the non-compactness of its gauge group, which prevents the application of the YM formalism {\it tout court}. Moreover, the delicate issue of matter couplings, which we addressed after writing down a dictionary to translate from pre-geometric to effective geometric quantities, arguably remains the least elegant result because of the complicated form of the effective kinetic terms. Nonetheless, these can be polynomial in the pre-geometric fields and reduce, after the phase transition, to the correct kinetic terms of quantum field theory on curved spacetime, via the correspondence principle. Notably, all terms of the Lagrangian density of the Standard Model of particles can be reconstructed in the unbroken phase with appropriate expressions.\\

Regarding the renormalisability of the effective kinetic terms for the matter fields in the unbroken phase, one can speculate about the possibility that vector fields, being the only ones that are potentially power-counting renormalisable, could act as a portal of mediators between the pre-geometric fields and the other matter fields. This could arise, for example, from integrating out heavy vector fields or from some non-perturbative dynamics like in non-linear sigma models. We postpone a more in-depth analysis of this aspect for now, leaving the door open to a possible unification of gravity and matter under a single gauge group which could be $SO(1,14)$, as this can be spontaneously broken first to $SO(1,4)\times SO(10)$ and then to $SO(1,3)\times SO(10)$, following the same line of reasoning adopted here. Another detailed study that is required to better understand this theory is the one concerning propagators and causality in a non-metric spacetime, with the latter admitting no straightforward description in the unbroken phase because of the absence of a metric.\\

As for the fluctuations of the pre-geometric fields which acquire mass as a result of the Higgs mechanism, that of the Higgs-like field depends on a free parameter and is expected to be around the Planck mass. This would be the most desirable possibility, given that then the phase transition due to the SSB would happen at about the Planck temperature. As a consequence, all alleged inconsistencies of classical gravity -- encountered in semi-classical arguments trying to merge its predictions with those of the quantum theory -- would be prevented from being realised in reality. The Planck temperature, in fact, would correspond to the critical temperature at which the fundamental gauge symmetry of spacetime is restored and the geometric description of gravity ceases to be meaningful. As for the fluctuations of the components of the gauge field $A_\mu^{AB}$, those corresponding to the unbroken generators of the SSB remain massless and are somehow the remnants of the pre-geometric YM theory in the spontaneously broken phase. On the other hand, those corresponding to the broken generators of the SSB gain a relatively small bare mass, but their effective mass ends up being much higher than that -- of the order of the Planck scale -- because of their interactions with gravity. Speaking of which, the possible existence of tachyons in the theory, which would seem to follow from the basic requirement of local Lorentz invariance, is not yet fully understood and shall be investigated better in the future. Some of the most interesting possibilities in this regard involve the decay of tachyons in the form of gravitational Cherenkov radiation \cite{lapedes:tachyons,schwartz:tachyons} and a consistent covariant quantisation of tachyonic fields (see Refs.\ \cite{paczos:tachyons,Jodlowski:2024rut} and references therein), which could have applications for the study of spacetime singularities and energy conditions. Regarding the Planckian tachyons obtained in our case, they will decay into gravitational Cherenkov radiation in a few Planckian times or so, in agreement with the estimates of Refs.\ \cite{lapedes:tachyons,schwartz:tachyons}. Hence they contribute to quantum loop effects, but cannot survive as asymptotic states.\\

Conceptually, a crucial step in our construction was the splitting, after the SSB, of all the effective geometric quantities into classical and quantum parts. The latter ones are determined by the surviving fluctuations of the pre-geometric fields, while the former give rise to the usual geometric quantities of classical gravity, including the spacetime metric and the spin connection. Within this framework, gravity can be interpreted, in a certain sense, as being `contained' in the pre-geometric fields of the non-metric spacetime and resulting, after the SSB of the fundamental gauge symmetry, from the classical values of those fields. This phase transition from a pre-geometric to a geometric, from a non-metric to a gravitational, from a quantum to a classical Universe brings about the notion of causality too, as such principle hinges on the existence of a spacetime metric. At energy scales near that one of the phase transition, one can expect the departures from the predictions of GR to become significant. Indeed, one of them is represented by the self-interaction terms for the fluctuations of the gauge field $A_\mu^{AB}$: these cannot be expressed in the standard metric, background-independent form of quantum field theory on curved spacetime; yet, their interaction strength is set at the Planck scale by the gravitational sector, thus fluctuations become relevant only in the high-energy regime. On the other hand, at low energies the theory effortlessly reproduces all results of GR, including gravitational waves with two possible polarisations. Even so, for now it remains unclear whether the associated massless spin-$2$ particle, i.e.\ the graviton, needs be quantised in this framework, given that it results from the classical values of the pre-geometric fields and that such description of gravity is phased-out once the fundamental gauge symmetry of spacetime is restored above the critical temperature in the high-energy limit.\\

More generally, we defer a proper discussion of the quantisation of the theory to upcoming works, which could address this question with a variety of methods: perturbation theory, given that we have stated the conditions under which the theory is potentially power-counting renormalisable in the gravitational sector; renormalisation group flow, given that the gravitational constant would be expected to run with the energy scale (according to the running of the coupling constant $k_\textup{W}$) and could possibly have nontrivial fixed points in the high-energy limit; functional integration and the path integral approach to compute correlators of fields; background-independent schemes, given that the theory is independent of the specifics of the pre-geometric spacetime in the unbroken phase.\\

In Sec.\ \ref{sec:10} we have presented the most straightforward extension of the theory that is based on the gauge group $SO(2,4)$, making use of two different multiplets of pre-geometric scalar fields -- one in the fundamental representation of $SO(1,4)$ and the other one in the fundamental representation of $SO(2,4)$ -- in order to realise two steps of SSB, first from $SO(2,4)$ to $SO(1,4)$ and then from $SO(1,4)$ to $SO(1,3)$. The analysis of the Higgs mechanism in this case shows that the additional Higgs-like field, which is introduced for the SSB of the conformal symmetry of spacetime, is a ghost scalar field. The presence of a ghost in the spectrum of this extended theory is somehow to be expected (in analogy to what has been found, for example, in the Refs.\ \cite{'t Hooft:conformal,mannheim:conformal}), even though it remains highly model-dependent in our framework. For this reason, future studies could lead perhaps to different outcomes.

\section{Conclusions and perspectives}\label{sec:12}
In this paper we have elaborated on a possible UV completion of GR inspired by a seminal work by {\it Wilczek} \cite{wilczek:gauge}. The SSB of a fundamental gauge symmetry of a pre-geometric four-dimensional spacetime can be described as part of the dynamics of a field theory {\it à la} Yang--Mills, in a manifestly generally covariant formulation. This can be achieved by introducing, other than a $SO(1,4)$ or $SO(3,2)$ gauge field $A_\mu^{AB}$, a Higgs-like field $\phi^A$ whose symmetry-breaking potential reduces the gauge symmetry of spacetime to that of the Lorentz group $SO(1,3)$. The pre-geometric spacetime of the unbroken phase is assumed to have no spacetime metric nor tetrads, with such geometric quantities of classical gravity arising only in the spontaneously broken phase after a natural identification of the components of $A_\mu^{AB}$. This allows to obtain the Einstein--Cartan theory in the low-energy limit and, thus, it provides meaning to the notions of spacetime curvature and gravity, starting from a gauge theory that would otherwise be unrelated to gravity.\\
 
Diffeomorphism invariance and the EEP are shown to be emergent after the SSB as well, with the validity of the latter principle being possible only in the classical limit where the quantum fluctuations of the pre-geometric fields $A_\mu^{AB}$ and $\phi^A$ can be neglected. Likewise, the two mass scales that characterise the classical theory of gravity, i.e.\ the Planck mass and the cosmological constant, can be identified only after the SSB as combinations of the free parameters of the theory. Our computations then yield the correct sign for the cosmological constant depending on whether the fundamental gauge group is taken to be either the de Sitter or the anti-de Sitter group. Therefore, everything related to a metric theory of gravity is emergent in this construction, from the physical quantities to their dynamical laws, from the fundamental principles to the energy scales, effectively arising from the intertwined dynamics of a gauge field and a Higgs-like field in a pre-geometric spacetime with no notion of metric distance or spacetime curvature.\\

The strict rigidity imposed in the construction of Lagrangian densities by the few allowed geometric ingredients of this picture, other than the pre-geometric fields $A_\mu^{AB}$ and $\phi^A$, means that $\mathcal{L}_\textup{W}$ is the \emph{only} physically viable theory of emergent gravity for a four-dimensional spacetime in the unbroken phase. Taking this into account, it is remarkable that it reduces, via a mechanism of SSB, to nothing else but classical gravity. Putting it in more emphatic terms, under the rather generic hypotheses considered in this work, a mechanism of SSB in a pre-geometric gauge theory leads inevitably to the emergence of a classical metric theory of gravity. This result highlights a profound link between gravity and the gauge principle applied to the description of spacetime.\\

The prospect of the fundamental laws of gravity changing at very high energies due to the restoration of a gauge symmetry of spacetime is ultimately the most fascinating possibility offered by this theory of emergent gravity, for it encourages alternative approaches to the quest of finding a UV completion of Einstein gravity. In the various frameworks of the Quantum Gravity research program, the most common outlook is to `quantise' gravity, in the sense of starting from the classical theory of gravity and applying the rules of the quantum theory to it. Instead, here we have attempted to do the opposite, that is `gravitationalise' the quantum, in the sense of starting from the rules of the quantum theory and letting gravity emerge. More than a quantum theory of gravity, this is in fact a gravitational theory of the quantum Universe.\\
\\

{\bf Acknowledgments}.
We would like to thank Francesco Cianfrani and Tomi Koivisto for valuable discussions and remarks on these subjects. AA acknowledges the support of the National Science Foundation of China (NSFC) through the grant No.\ 12350410358; the Talent Scientific Research Program of College of Physics, Sichuan University, Grant No.\ 1082204112427; the Fostering Program in Disciplines Possessing Novel Features for Natural Science of Sichuan University, Grant No.\ 2020SCUNL209 and 1000 Talent program of Sichuan province 2021. SC and GM acknowledge the support of Istituto Nazionale di Fisica Nucleare, Sez.\ di Napoli, Iniziative Specifiche QGSKY and MoonLight-2, Italy. AM acknowledges the support by the NSFC, through the grant No.\ 11875113, the Shanghai Municipality, through the grant No.\ KBH1512299, and by Fudan University, through the grant No.\ JJH1512105.

\appendix
\noappendicestocpagenum
\addappheadtotoc
\section{Spontaneous symmetry breaking mechanism}\label{sec:appendix_A}
\setcounter{equation}{0}\renewcommand{\theequation}{A\arabic{equation}}
In this appendix we show explicitly the mechanism of SSB for both theories $\mathcal{L}_\textup{MM}$ and $\mathcal{L}_\textup{W}$. According to the Eq.\ \eqref{eq:SSB_MM}, for $\mathcal{L}_\textup{MM}$ we find that
\begin{equation}
    \begin{split}
    \mathcal{L}_\textup{MM}&=k_\textup{MM}\epsilon_{ABCDE}\epsilon^{\mu\nu\rho\sigma}F_{\mu\nu}^{AB}F_{\rho\sigma}^{CD}\phi^E\xrightarrow{SSB}k_\textup{MM}v(\epsilon_{abcd}\epsilon^{\mu\nu\rho\sigma}R_{\mu\nu}^{ab}R_{\rho\sigma}^{cd}\\
    &\mp4m^2\epsilon_{abcd}\epsilon^{\mu\nu\rho\sigma}e_\mu^ae_\nu^bR_{\rho\sigma}^{cd}+4m^4\epsilon_{abcd}\epsilon^{\mu\nu\rho\sigma}e_\mu^ae_\nu^be_\rho^ce_\sigma^d).
    \end{split}
\end{equation}

To compute the first of these three terms, one can take into account what follows. The Lorentzian signature of the Minkowski metric implies that $\epsilon_{abcd}=-\epsilon^{abcd}$. Using the following linear algebra identity for an $n\times n$ matrix $M_{ij}$,
\begin{equation}
    \epsilon^{j_1\dots j_n}M_{i_1j_1}\dots M_{i_nj_n}=\det{M}\epsilon^{i_1\dots i_n},
\end{equation}
from which it follows that
\begin{equation}
    \epsilon^{abcd}e_a^\mu e_b^\nu e_c^\rho e_d^\sigma=\frac{1}{e}\epsilon^{\mu\nu\rho\sigma},
\end{equation}
one finds that
\begin{equation}
    \begin{split}
    \epsilon_{abcd}\epsilon^{\mu\nu\rho\sigma}&=e\epsilon_{abcd}\epsilon^{ijkl}e_i^\mu e_j^\nu e_k^\rho e_l^\sigma\\
    &=-e\begin{vmatrix}
        \delta_a^i&\delta_b^i&\delta_c^i&\delta_d^i\\
        \delta_a^j&\delta_b^j&\delta_c^j&\delta_d^j\\
        \delta_a^k&\delta_b^k&\delta_c^k&\delta_d^k\\
        \delta_a^l&\delta_b^l&\delta_c^l&\delta_d^l
    \end{vmatrix}e_i^\mu e_j^\nu e_k^\rho e_l^\sigma=-e\begin{vmatrix}
        e_a^\mu&e_b^\mu&e_c^\mu&e_d^\mu\\
        e_a^\nu&e_b^\nu&e_c^\nu&e_d^\nu\\
        e_a^\rho&e_b^\rho&e_c^\rho&e_d^\rho\\
        e_a^\sigma&e_b^\sigma&e_c^\sigma&e_d^\sigma
    \end{vmatrix},
    \end{split}
\end{equation}
where it was deployed the identity
\begin{equation}
    \epsilon_{abcd}\epsilon^{ijkl}=-\begin{vmatrix}
        \delta_a^i&\delta_b^i&\delta_c^i&\delta_d^i\\
        \delta_a^j&\delta_b^j&\delta_c^j&\delta_d^j\\
        \delta_a^k&\delta_b^k&\delta_c^k&\delta_d^k\\
        \delta_a^l&\delta_b^l&\delta_c^l&\delta_d^l
    \end{vmatrix}.
\end{equation}
The first term of $\mathcal{L}_\textup{MM}$ can now be shown to be proportional to the Gauss--Bonnet term, with
\begin{equation}\label{eq:gauss-bonnet}
    \mathcal{G}\equiv R^2-4R_{\mu\nu}R^{\mu\nu}+R_{\mu\nu\rho\sigma}R^{\mu\nu\rho\sigma}=(e_a^\mu e_b^\nu e_c^\rho e_d^\sigma-4e_a^\mu e_d^\nu e_c^\rho e_b^\sigma+e_c^\mu e_d^\nu e_a^\rho e_b^\sigma)R_{\mu\nu}^{ab}R_{\rho\sigma}^{cd},
\end{equation}
and so it does not contribute to the equations of motion if the spacetime manifold is four-dimensional, namely 
\begin{equation}
    \begin{split}
    \epsilon_{abcd}\epsilon^{\mu\nu\rho\sigma}R_{\mu\nu}^{ab}R_{\rho\sigma}^{cd}&=-e\begin{vmatrix}
        e_a^\mu&e_b^\mu&e_c^\mu&e_d^\mu\\
        e_a^\nu&e_b^\nu&e_c^\nu&e_d^\nu\\
        e_a^\rho&e_b^\rho&e_c^\rho&e_d^\rho\\
        e_a^\sigma&e_b^\sigma&e_c^\sigma&e_d^\sigma
    \end{vmatrix}R_{\mu\nu}^{ab}R_{\rho\sigma}^{cd}\\
    &=-e(4R^2-16R_{\mu\nu}R^{\mu\nu}+4R_{\mu\nu\rho\sigma}R^{\mu\nu\rho\sigma})=-4e\mathcal{G}.
    \end{split}
\end{equation}

Assuming the $4\times4$ matrix of the gauge potentials $A_\mu^{a5}$ to be invertible, it can be shown that
\begin{equation}\label{eq:id_EH}
    \epsilon_{abcd}\epsilon^{\mu\nu\rho\sigma}e_\mu^ae_\nu^bR_{\rho\sigma}^{cd}=-4ee_a^\mu e_b^\nu R_{\mu\nu}^{ab},
\end{equation}
hence the second term of $\mathcal{L}_\textup{MM}$ is proportional to $\mathcal{L}_\textup{EH}$. To prove this result, recall that the Levi-Civita tensor $\varepsilon$ is related to the Levi-Civita symbol $\epsilon$ by the relations \cite{carroll:spacetime}
\begin{equation}
    \varepsilon_{\mu\nu\rho\sigma}=e\epsilon_{\mu\nu\rho\sigma},\qquad\varepsilon^{\mu\nu\rho\sigma}=\frac{1}{e}\epsilon^{\mu\nu\rho\sigma};
\end{equation}
in particular, $\varepsilon_{abcd}=\epsilon_{abcd}$ in the case of the Minkowski metric. The difference between the Levi-Civita symbol and the corresponding tensor is important because the latter, being a proper tensor rather than a tensor density, can be expressed with indices which can be raised and lowered by means of a metric. That said, one can compute that
\begin{equation}
    \begin{split}
        \epsilon_{abcd}\epsilon^{\mu\nu\rho\sigma}e_\mu^ae_\nu^bR_{\rho\sigma}^{cd}&=e\varepsilon_{abcd}\varepsilon^{\mu\nu\rho\sigma}e_\mu^ae_\nu^bR_{\rho\sigma}^{cd}=e\varepsilon_{\mu\nu cd}\varepsilon^{\mu\nu\rho\sigma}R_{\rho\sigma}^{cd}=e\varepsilon_{\mu\nu\tau\xi}\varepsilon^{\mu\nu\rho\sigma}e^\tau_ce^\xi_dR_{\rho\sigma}^{cd}\\
        &=-4e\delta^\rho_{[\tau}\delta^\sigma_{\xi]}e^\tau_ce^\xi_dR_{\rho\sigma}^{cd}=-4ee^{[\rho}_ce^{\sigma]}_dR_{\rho\sigma}^{cd}=-4ee_a^\mu e_b^\nu R_{\mu\nu}^{ab},
    \end{split}
\end{equation}
where it was made use of the identity
\begin{equation}
    \varepsilon_{\mu\nu\tau\xi}\varepsilon^{\mu\nu\rho\sigma}=-4\delta^\rho_{[\tau}\delta^\sigma_{\xi]}.
\end{equation}

Finally, using the following linear algebra identity for an $n\times n$ matrix $M_{ij}$,
\begin{equation}
    \det{M}=\frac{1}{n!}\epsilon^{i_1\dots i_n}\epsilon^{j_1\dots j_n}M_{i_1j_1}\dots M_{i_nj_n},
\end{equation}
one finds that
\begin{equation}\label{eq:id_cc}
    \epsilon_{abcd}\epsilon^{\mu\nu\rho\sigma}e_\mu^a e_\nu^b e_\rho^c e_\sigma^d=-\epsilon^{abcd}\epsilon^{\mu\nu\rho\sigma}e_\mu^a e_\nu^b e_\rho^c e_\sigma^d=-24e,
\end{equation}
thus the third term of $\mathcal{L}_\textup{MM}$ is a cosmological constant term.\\

Having completed the computations for $\mathcal{L}_\textup{MM}$, it is now straightforward to perform analogous computations for $\mathcal{L}_\textup{W}$, which yield
\begin{equation}
    \begin{split}
    \mathcal{L}_\textup{W}&=k_\textup{W}\epsilon_{ABCDE}\epsilon^{\mu\nu\rho\sigma}F_{\mu\nu}^{AB}\nabla_\rho\phi^C\nabla_\sigma\phi^D\phi^E\xrightarrow{SSB}k_\textup{W}v^3\epsilon_{ABCDE}\epsilon^{\mu\nu\rho\sigma}F_{\mu\nu}^{AB}A_\rho^{C5}A_\sigma^{D5}\delta^E_5\\
    &=k_\textup{W}v^3m^2\epsilon_{abcd}\epsilon^{\mu\nu\rho\sigma}(R_{\mu\nu}^{ab}\mp2m^2e_{[\mu}^ae_{\nu]}^b)e_\rho^ce_\sigma^d\\
    &=k_\textup{W}v^3m^2(\epsilon_{abcd}\epsilon^{\mu\nu\rho\sigma}e_\rho^ce_\sigma^dR_{\mu\nu}^{ab}\mp2m^2\epsilon_{abcd}\epsilon^{\mu\nu\rho\sigma}e_\mu^ae_\nu^be_\rho^ce_\sigma^d)\\
    &=-4k_\textup{W}v^3m^2ee^\mu_ae^\nu_bR_{\mu\nu}^{ab}\pm48k_\textup{W}v^3m^4e,
    \end{split}
\end{equation}
where in the last step Eqs.\ \eqref{eq:id_EH} and \eqref{eq:id_cc} were deployed.

\section{Higgs mechanism}\label{sec:appendix_B}
\setcounter{equation}{0}\renewcommand{\theequation}{B\arabic{equation}}
This appendix is devoted to showing the Higgs mechanism for the Lagrangian density \eqref{eq:lagrangian} in detail. All the calculations presented here are carried out up to the second order in the fluctuations of the pre-geometric fields. To begin with, the expression \eqref{eq:cov_dev} for the internal-space covariant derivative acting on internal-space vectors needs be updated as follows:
\begin{equation}
    \nabla_\mu\phi^A\xrightarrow{SSB}\delta_5^A\partial_\mu\rho\pm(vme_\mu^a+v\sigma_\mu^a+me_\mu^a\rho+\sigma_\mu^a\rho).
\end{equation}
With this expression, the effective Jacobian determinant \eqref{eq:aux_J} becomes
\begin{equation}
    \begin{split}
    J\xrightarrow{SSB}&-24v^5m^4e+4v^4m^3(v\epsilon_{abcd}\epsilon^{\mu\nu\rho\sigma}e_\mu^ae_\nu^be_\rho^c\sigma_\sigma^d-30me\rho)\\
    &+2v^3m^2(3v^2\epsilon_{abcd}\epsilon^{\mu\nu\rho\sigma}e_\mu^ae_\nu^b\sigma_\rho^c\sigma_\sigma^d+10vm\epsilon_{abcd}\epsilon^{\mu\nu\rho\sigma}e_\mu^ae_\nu^be_\rho^c\sigma_\sigma^d\rho-120m^2e\rho^2).
    \end{split}
\end{equation}

We can now proceed to the expansion of the relevant terms in the total Lagrangian density. For the gravitational sector, one may derive
\begin{equation}
    \begin{split}
    \mathcal{L}_\textup{MM}\xrightarrow{SSB}&(\pm16k_\textup{MM}vm^2ee_a^\mu e_b^\nu R_{\mu\nu}^{ab}-96k_\textup{MM}vm^4e-4k_\textup{MM}ve\mathcal{G})\\
    &[(\pm16k_\textup{MM}m^2ee_a^\mu e_b^\nu R_{\mu\nu}^{ab}-96k_\textup{MM}m^4e-4k_\textup{MM}e\mathcal{G})\rho\\
    &+4k_\textup{MM}v\epsilon_{abcd}\epsilon^{\mu\nu\rho\sigma}(R_{\mu\nu}^{ab}\mp2m^2e_\mu^ae_\nu^b)(\partial_\rho\sigma_\sigma^{cd}+2\omega_\sigma^{fd}\sigma_{f\rho}^c\mp2me_\rho^c\sigma_\sigma^d)]\\
    &+4k_\textup{MM}\epsilon_{abcd}\epsilon^{\mu\nu\rho\sigma}[(R_{\mu\nu}^{ab}\mp2m^2e_\mu^ae_\nu^b)(\partial_\rho\sigma_\sigma^{cd}\rho+2\omega_\sigma^{fd}\sigma_{f\rho}^c\rho\mp2me_\rho^c\sigma_\sigma^d\rho)\\
    &+v(\partial_\mu\sigma_\nu^{ab}\partial_\rho\sigma_\sigma^{cd}+2\partial_\mu\omega_\nu^{ab}\sigma_{f\rho}^c\sigma_\sigma^{fd}\mp2\partial_\mu\omega_\nu^{ab}\sigma_\rho^c\sigma_\sigma^d+4\omega_\sigma^{fd}\partial_\mu\sigma_\nu^{ab}\sigma_{f\rho}^c\\
    &\mp4me_\rho^c\partial_\mu\sigma_\nu^{ab}\sigma_\sigma^d+6\omega_\nu^{eb}\omega_\sigma^{fd}\sigma_{e\mu}^a\sigma_{f\rho}^c\mp2\omega_{e\mu}^a\omega_\nu^{eb}\sigma_\rho^c\sigma_\sigma^d\mp8m\omega_\nu^{eb}e_\rho^c\sigma_{e\mu}^a\sigma_\sigma^d\\
    &\mp2m^2e_\rho^ce_\sigma^d\sigma_{e\mu}^a\sigma_\nu^{eb}+6m^2e_\mu^ae_\nu^b\sigma_\rho^c\sigma_\sigma^d)]
    \end{split}
\end{equation}
and
\begin{equation}\label{eq:L_W-SSB}
    \begin{split}
    \mathcal{L}_\textup{W}\xrightarrow{SSB}&(-4k_\textup{W}v^3m^2ee^\mu_ae^\nu_bR_{\mu\nu}^{ab}\pm48k_\textup{W}v^3m^4e)\\
    &+[3(-4k_\textup{W}v^2m^2ee^\mu_ae^\nu_bR_{\mu\nu}^{ab}\pm48k_\textup{W}v^2m^4e)\rho+2k_\textup{W}v^2m\epsilon_{abcd}\epsilon^{\mu\nu\rho\sigma}(vme_\rho^ce_\sigma^d\partial_\mu\sigma_\nu^{ab}\\
    &+ve_\rho^cR_{\mu\nu}^{ab}\sigma_\sigma^d\mp4vm^2e_\mu^ae_\nu^be_\rho^c\sigma_\sigma^d+2vme_\rho^ce_\sigma^d\omega_\nu^{eb}\sigma_{e\mu}^a)]\\
    &+k_\textup{W}v\epsilon_{abcd}\epsilon^{\mu\nu\rho\sigma}[R_{\mu\nu}^{ab}(v^2\sigma_\rho^c\sigma_\sigma^d+4vme_\rho^c\sigma_\sigma^d\rho+3m^2e_\rho^ce_\sigma^d\rho^2)\\
    &+2vm(2ve_\rho^c\partial_\mu\sigma_\nu^{ab}\sigma_\sigma^d+3me_\rho^ce_\sigma^d\partial_\mu\sigma_\nu^{ab}\rho+vme_\rho^ce_\sigma^d\sigma_{e\mu}^a\sigma_\nu^{eb}+4v\omega_\nu^{eb}e_\rho^c\sigma_{e\mu}^a\sigma_\sigma^d\\
    &+6m\omega_\nu^{eb}e_\rho^ce_\sigma^d\sigma_{e\mu}^a\rho\mp6vme_\mu^ae_\nu^b\sigma_\rho^c\sigma_\sigma^d\mp16m^2e_\mu^ae_\nu^be_\rho^c\sigma_\sigma^d\rho)\mp6m^4e_\mu^ae_\nu^be_\rho^ce_\sigma^d\rho^2],
    \end{split}
\end{equation}
where the $0$\textsuperscript{th}, the $1$\textsuperscript{st} and the $2$\textsuperscript{nd} order terms have been properly grouped.\\

The symmetry-breaking potential shows no $0$\textsuperscript{th} nor $1$\textsuperscript{st} order contributions, and reduces to
\begin{equation}\label{eq:L_SSB-SSB}
    \mathcal{L}_\textup{SSB}\xrightarrow{SSB}-96k_\textup{SSB}v^3m^4e\rho^2.
\end{equation}

Finally, the effective kinetic term for the pre-geometric field $\phi^A$ yields
\begin{equation}
    \begin{split}
    \mathcal{L}_\phi\xrightarrow{SSB}&-\frac{1}{2}m^2\sqrt{-g}(\pm g^{\mu\nu}\partial_\mu\rho\partial_\nu\rho+4v^2m^2+2v^2mg^{\mu\nu}\eta_{ab}e_\mu^a\sigma_\nu^b+8vm^2\rho\\
    &+v^2g^{\mu\nu}\eta_{ab}\sigma_\mu^a\sigma_\nu^b+4vmg^{\mu\nu}\eta_{ab}e_\mu^a\sigma_\nu^b\rho+4m^2\rho^2),
    \end{split}
\end{equation}
where the $1$\textsuperscript{st} and $2$\textsuperscript{nd} order terms arising from the expansion of the effective geometric quantity $J^{-1}w_A^\mu w_B^\nu\eta^{AB}$ beyond the $0$\textsuperscript{th} order were omitted for conciseness.

\section{Computations for the conformal symmetry}\label{sec:appendix_C}
\setcounter{equation}{0}\renewcommand{\theequation}{C\arabic{equation}}
In this appendix we illustrate the relevant calculations for the Higgs mechanism of a theory of emergent gravity based on the gauge symmetry of the group $SO(2,4)$, as described in Sec.\ \ref{sec:10}. Once again, calculations are carried out up to the second order in the fluctuations of the pre-geometric fields. After the $\widetilde{\text{SSB}}$ from $SO(2,4)$ to $SO(1,4)$, the expression \eqref{eq:cov_dev} for the internal-space covariant derivative acting on internal-space vectors yields:
\begin{equation}
    \nabla_\mu\Tilde{\phi}^{\Tilde{A}}\xrightarrow{\widetilde{SSB}}\delta_2^{\Tilde{A}}\partial_\mu\Tilde{\rho}-A_\mu^{\Tilde{A}2}(\Tilde{v}+\Tilde{\rho}).
\end{equation}

For the gravitational sector one finds that
\begin{equation}
    \mathcal{L}_\textup{W}\xrightarrow{\widetilde{SSB}}k_\textup{W}\epsilon_{ABCDE}\epsilon^{\mu\nu\rho\sigma}F_{\mu\nu}^{AB}\nabla_\rho\phi^C\nabla_\sigma\phi^D\phi^E(1+\Tilde{\rho}/\Tilde{v}),
\end{equation}
with the ensuing SSB that leads to the same expression as in Eq.\ \eqref{eq:L_W-SSB}, times the factor $(1+\Tilde{\rho}/\Tilde{v})$.\\

The symmetry-breaking potential reduces to
\begin{equation}
    \mathcal{L}_{\widetilde{\textup{SSB}}}\xrightarrow{\widetilde{SSB}}-4k_{\widetilde{\textup{SSB}}}\lvert J\rvert(\Tilde{\rho}/\Tilde{v})^2\xrightarrow{SSB}-96k_{\widetilde{\textup{SSB}}}v^5m^4e(\Tilde{\rho}/\Tilde{v})^2,
\end{equation}
while the effective kinetic term of the pre-geometric field $\Tilde{\phi}^{\Tilde{A}}$ yields
\begin{equation}
    \begin{split}
    \mathcal{L}_{\Tilde{\phi}}&\xrightarrow{\widetilde{SSB}}\frac{1}{3}v^{-3}J^{-1}w_A^\mu w_B^\nu\eta^{AB}[-\partial_\mu\Tilde{\rho}\partial_\nu\Tilde{\rho}+\eta_{\Tilde{C}\Tilde{D}}A_\mu^{\Tilde{C}2}A_\nu^{\Tilde{D}2}(\Tilde{v}+\Tilde{\rho})^2]\\
        &\xrightarrow{SSB}-\frac{1}{2}m^2\sqrt{-g}g^{\mu\nu}[-\partial_\mu\Tilde{\rho}\partial_\nu\Tilde{\rho}+\eta_{\Tilde{C}\Tilde{D}}A_\mu^{\Tilde{C}2}A_\nu^{\Tilde{D}2}(\Tilde{v}+\Tilde{\rho})^2],
    \end{split}
\end{equation}
where, once again, the $1$\textsuperscript{st} and the $2$\textsuperscript{nd} order terms originated from the expansion of the effective geometric quantity $J^{-1}w_A^\mu w_B^\nu\eta^{AB}$, beyond the $0$\textsuperscript{th} order, were omitted for conciseness.

\end{document}